\documentclass[11pt]{article}

\usepackage[scaled=0.92]{helvet}	% set Helvetica as the sans-serif font
\usepackage{amsmath}
\usepackage{amssymb} %conflicts with mtpro2
\usepackage{mathrsfs}

\usepackage{graphicx}

\usepackage{float}
\usepackage{subcaption}

\usepackage[dvipsnames]{xcolor}
\usepackage{hyperref}
\hypersetup{
    colorlinks=true,
    linkcolor=MidnightBlue,   
    filecolor=MidnightBlue,
    urlcolor=MidnightBlue,     
    citecolor=MidnightBlue,
    pdfsubject={Mathematics},
    bookmarksnumbered=true,
    bookmarksopen=true,
    bookmarksopenlevel=1,
    pdfstartview=Fit,
    pdfpagemode=UseOutlines,
}

\newtheorem{definition}{Definition}

\newtheorem{example}{Example}

\newtheorem{assumption}{Assumption}

\def\qed{ \ \vrule width.2cm height.2cm depth0cm\smallskip}
\newenvironment{proof}{\noindent {\bf Proof.\/}}{$\qed$\vskip 0.1in}

\newcommand{\ind}[1]{{\bf 1}_{\left\{ {#1} \right\}} }

\newcommand{\nob}{\noindent$\bullet$\ }

\newcommand{\ba}{\begin{array}}
\newcommand{\ea}{\end{array}}
\newcommand{\be}{\begin{equation}}
\newcommand{\ee}{\end{equation}}
\newcommand{\bea}{\begin{eqnarray}}
\newcommand{\eea}{\end{eqnarray}}
\newcommand{\beaa}{\begin{eqnarray*}}
\newcommand{\eeaa}{\end{eqnarray*}}

\def\dbA{\mathbb{A}}

\def\dbE{\mathbb{E}}
\def\dbF{\mathbb{F}}

\def\dbN{\mathbb{N}}
\def\dbP{\mathbb{P}}
\def\dbR{\mathbb{R}}
\def\dbS{\mathbb{S}}
\def\dbT{\mathbb{T}}

\def\dbX{\mathbb{X}}

\def\sE{\mathscr{E}}
\def\sM{\mathscr{M}}

\def\sA{\mathscr{A}}

\def\fL{\mathfrak{L}}

\def\a{\alpha}

\def\d{\delta}
\def\e{\varepsilon}

\def\k{\kappa}

\def\o{\omega}

\def\G{\Gamma}

\def\O{\Omega}
\def\cA{{\cal A}}
\def\cB{{\cal B}}

\def\cE{{\cal E}}
\def\cF{{\cal F}}

\def\cL{{\cal L}}

\def\cN{{\cal N}}
\def\cO{{\cal O}}
\def\cP{{\cal P}}

\def\qed{ \hfill \vrule width.25cm height.25cm depth0cm\smallskip}

\newcommand{\basa}{\begin{assumption}}
\newcommand{\easa}{\end{assumption}}

\newcommand{\bas}{\begin{assum}}
\newcommand{\eas}{\end{assum}}

\def\limsup{\mathop{\overline{\rm lim}}}
\def\liminf{\mathop{\underline{\rm lim}}}

\def\ba{{\bf a}}

\def\1{{\bf 1}}

\def\:{\!:\!}

\def\U{{\Upsilon}}

at 9pt

\newtheorem{thm}{Theorem}[section]

\newtheorem{assum}[thm]{Assumption}

\numberwithin{equation}{section}

\DeclareMathOperator*{\argmax}{arg\,max}
\renewcommand{\limsup}{\operatorname*{limsup}}
\renewcommand{\liminf}{\operatorname*{liminf}}

\newcommand{\con}{\mathrm{con}}
\newcommand{\obj}{{\mathrm{obj}}}
\newcommand{\rel}{{\mathrm{rel}}}
\newcommand{\pre}{\mathrm{pre}}

\begin{document}

\title{The Learning Approach to Games}
\author{Melih \.{I}\c{s}eri\footnote{Department of Mathematics,
    University of Michigan, United States, iseri@umich.edu.}\quad
  Erhan Bayraktar\footnote{Department of Mathematics, University of
    Michigan, United States, erhan@umich.edu.}}

\maketitle

\begin{abstract}
  This work introduces a unified framework for analyzing games in
  greater depth. In the existing literature, players’ strategies are
  typically assigned scalar values, and equilibrium concepts are used
  to identify compatible choices. However, this approach neglects the
  internal structure of players, thereby failing to accurately model
  observed behaviors.
  
  To address this limitation, we propose an abstract definition of a
  player, consistent with constructions in reinforcement learning.
  Instead of defining games as external settings, our framework
  defines them in terms of the players themselves. This offers a
  language that enables a deeper connection between games and
  learning. To illustrate the need for this generality, we study a
  simple two-player game and show that even in basic settings, a
  sophisticated player may adopt dynamic strategies that cannot be
  captured by simpler models or compatibility analysis.

  For a general definition of a player, we discuss natural conditions
  on its components and define competition through their behavior. In
  the discrete setting, we consider players whose estimates largely
  follow the standard framework from the literature. We explore
  connections to correlated equilibrium and highlight that dynamic
  programming naturally applies to all estimates. In the mean-field
  setting, we exploit symmetry to construct explicit examples of
  equilibria. Finally, we conclude by examining relations to
  reinforcement learning.
\end{abstract}


\newpage

\section{Introduction}\label{sec:introduction}

Game theory, like every branch of mathematics, explores fundamental
concepts for systems that are as general as possible, where individual
components have potential choices to make. The opportunities for
exploration are vast and deeply complex, with implications across a
diverse array of fields. These include societal structures,
competitions in numerous games, various dynamics of businesses,
computational decision processes, financial models, and many
more. Recent advancements have even surpassed human capabilities in
various domains, prompting increased efforts to better understand our
brains, the most fascinating dynamic system.

To motivate our goal intuitively, consider a swarm of self-driving
cars operating in a city. From the traditional game-theoretic
viewpoint, one might analyze the density of cars on the streets and
emerging traffic patterns. For such a problem, it is neither feasible
nor meaningful to account for the detailed internal structure of each
vehicle. As a result, however, it cannot be used to specify or
simulate how the cars themselves make decisions. We remark that such
traditional methods remain rooted in a central-planner
perspective. Roughly speaking, equilibrium concepts involve solving a
control problem, augmented by a compatibility condition among agents’
strategies. When the goal is to design optimal policies within
well-defined environments, the traditional framework remains
appropriate. However, when the focus shifts to designing complex
players themselves, such as those enabled by modern reinforcement
learning, the available methods are powerful, but a unifying language
is lacking. This work aims to fill this gap.

The main approach of this work is to separate the concept of player
from the external settings we design. This will provide a broader
perspective on games, in which multiple complex decision-makers
interact. In the literature, when there is a single decision-maker,
designing external rewards in the environment guides which future
observations the player will prefer. This, in turn, provides strong
guidance for designing players that act in our interests. Still,
designing players to tackle real-world problems requires significant
effort beyond setting up the environment. We emphasize this
perspective because once there is more than one player, the
expressiveness of setting up the environment diminishes. We argue that
omitting the structure of players and considering only stability
conditions over the set of strategies cannot provide a sufficiently
rich understanding.

Let us discuss two illustrative examples of competitive settings. The
first concerns one of the simplest competitive games,
rock-paper-scissors. Suppose a player competes against ten opponents
in a tournament and can perfectly generate a uniform distribution over
actions. Will the player choose the uniform equilibrium strategy? In
that case, winning the tournament is highly unlikely. Of course, an
equilibrium strategy is unexploitable, but players aiming to win the
tournament would naturally avoid it. Recognizing that no one would
actually use the uniform strategy, the player’s goal becomes twofold:
to estimate the opponents’ behavior and simultaneously to deceive
their estimates. This shows two simple but crucial points: (i) for
competing players, the objective is not to be unexploitable, and (ii)
the essence of the game lies not in its formal rules, but in the
players themselves. Even when the setting is simple, the strategic
interplay among players can exhibit profound complexity, revealing
that competition lies not in the game’s structure but in its
participants. The second example conveys the same ideas, but relies on
the structure of the game rather than human sensory
interactions. Suppose two players repeatedly change states in ${0,1}$,
with their actions representing transition probabilities. One player
aims to "catch" the other, while the second aims to "evade"; their
rewards are defined in a zero-sum fashion. The same question arises:
will they simply choose equal probabilities to remain unexploitable
and earn nothing on average? In reality, if the players are truly
competing, they will constantly change their strategies. For instance,
the evading player might begin appearing in state $1$ more frequently
to lure the opponent into following, only to deceive them for a
short-term gain. Note that no equilibrium strategy yields more than
zero expected gain for either side, yet that is not what the players
seek. The example motivates that competition is defined not by
convergence, but by the very effort to avoid it. Hence, in the next
section, we will say that players are cooperating when their behaviors
actually converge. As these examples show, such cooperation may occur
involuntarily, when one player seeks merely to remain unexploitable.

In Section \ref{sec:defining-players}, we introduce the main
definition of a player, and discuss some intrinsic objectives of their
components. In Section \ref{sec:Discrete-Games}, we explore discrete
games, treating mainly the standard components as estimates of the
player. After presenting specific estimates, we define uncertain
equilibrium to impose conditions such as optimality and recurrence. We
then refine this notion by adding further conditions such as
consistency and psychological states. This allows a richer
characterization of both players and the equilibrium concept. We also
demonstrate connections between our optimality assumption and the
concept of correlated equilibrium. Later, we present a toy two-player
game to illustrate the dynamic nature of games in the simplest
settings. In Section \ref{sec:stated-mean-field-games}, we examine
mean-field games with constant estimates, except that the
representative player estimates the population strategies. As
observations can be generated by relying on symmetries, we introduce a
learning algorithm with explicit uncertain equilibrium. In
Section~\ref{sec:Control}, we provide further connections to core
structures in reinforcement learning, present a basic learning
algorithm that does not rely on standard value-based methods, and
review related concepts in multi-agent reinforcement learning.

\emph{Some related literature.} \sloppy 
Classic non-cooperative game theory traces back to the von
Neumann-Morgenstern seminal work \cite{NM1944} and Nash
\cite{N1951}. We refer to the book Maschler-Solan-Zamir \cite{MSZ2013}
for a comprehensive treatment. There are many notions of equilibrium,
and we highlight the correlated equilibrium introduced by Aumann
\cite{A1987}. The author raises common criticisms of the classic Nash
equilibrium and introduces correlated equilibrium, which incorporates
randomness in players' strategies. A correlated equilibrium is a
single distribution over the players' strategies and is therefore
typically motivated by a mediator who draws from this
distribution. However, a single distribution cannot capture the
differing knowledge each player may possess. Because of this
structural connection, we will offer a comparison after introducing
the uncertain equilibrium for discrete games.

Critiques of treating equilibrium concepts as central to games appear
across several literatures. Kadane and Larkey \cite{KL1982}, for
example, adopt a subjective perspective, formalizing views that had
previously been discussed informally (as reflected in the Editor’s
Note) and bringing them into the management science framework. In the
multi-agent reinforcement learning literature, the technical note by
Shoham, Powers and Grenager \cite{SPG2003} (see also \cite{SPG2007})
criticizes the lack of conceptual clarity arising from the unjustified
use of equilibrium concepts as both learning objectives and evaluation
criteria. In the behavioral game theory literature, Wright and
Leyton-Brown \cite{WL2017} demonstrate that Nash equilibrium is a poor
description of human behavior even in simple games, and provide a
systematic comparison of different models of human behavior. We remark
that all such models still omit a model of the human itself, mapping
an external game description directly to a distribution over
actions. Hartford, Wright, and Leyton-Brown \cite{HWB2016} demonstrate
that a neural network, essentially a function approximation,
outperforms all such behavioral models. This strengthens the
motivation for this work, as we argue that the primary modeling object
should be the players themselves, and that games cannot be reduced to
their external descriptions.

Learning in games has a rich literature, beginning with Brown
\cite{B1951}, who introduced the notion of fictitious play. Players
are assumed to have predefined learning rules, and the question is
whether the long-run average of observed actions converges to an
equilibrium. Although such convergence is not always guaranteed (see,
for example, Daskalakis et al. \cite{DFPPV2010}), Hart and Mas-Colell
\cite{HM2000} combine regret with fictitious play to show convergence
to correlated equilibrium. For a comprehensive treatment, see the book
Fudenberg-Levine \cite{FL1998}.

To address games with a large number of players, where equilibria
become intractable, Lasry-Lions \cite{LL2007} and, independently,
Huang-Malham\'e-Caines \cite{HMC2006} introduced the concept of
mean-field games. Since then, the framework has been extensively
studied. In this setting, agents interact only through the empirical
distribution of their states and are indistinguishable from one
another, allowing a continuum limit to be identified with a
representative agent. We refer the reader to the excellent two-volume
book Carmona-Delarue \cite{CD2018a,CD2018b}.

In our framework, players may favor having a large collection of
estimates, each of which must be learned from observations. We cannot
hope to cover every relevant learning algorithm and its extensive
literature, however, we will highlight some connections in Section
\ref{sec:Control}. Here, we would like to point out the diverse work
on random value functions. We also advocate that the value is
fundamentally unknown to a learning player and is one of the main
sources of uncertainty in planning future behavior. We refer to
Thurstone \cite{T1927}, Luce \cite{L1959}, Block \cite{B1974},
McFadden \cite{M1974}, Train \cite{T2009}, and references therein for
discussions rooted in psychology and economics. One of the oldest
approaches is Thompson sampling, introduced in Thompson \cite{T1933}
and recently popularized by the empirical study Chapelle-Li
\cite{CL2011}. Upper confidence bound algorithms in the context of
multi-armed bandit problems, see for example Auer et
al. \cite{ACF2002}, can also be viewed as a random value approach.
Lastly, let us mention Bellemare-Dabney-Munos
\cite{bellemare2017distributional} who prominently promoted modeling
the value function as a random variable in the context of Markov
decision processes with applications to reinforcement learning.

\section{Definition of Players}
\label{sec:defining-players}

In this section, we provide a definition of a player and elaborate on
general requirements we may impose. At the end, we also define
cooperation and competition through their behavior. Let us first
introduce some preliminary definitions and notation:

\nob Universe, or the environment, is an abstract probability space
$(\O^u,\cF^u, \dbP^u$);

\nob $\cP(E)$ denotes the set of probability distributions on an
arbitrary set $E$;

\nob $\cE$ is the space of observables, and $\sE$ is the set of finite
sequences of $\cE$;

\nob $\dbA$ is the space of actions, and $\sA$ is the set of finite
sequences of $\dbA$;

\nob $\sM_\U$ is the set of functions taking values in $\dbA$, called
the space of behaviors;

\nob $\sM_\varphi$ is the set of functions, called the space of
estimates.

Similar to actions, we are keeping observables abstract. We have not
yet specified the domains and ranges of the estimations. Also,
$\varphi$ denotes an index for estimates. We will introduce a
collection of them in the upcoming sections.

In the realm of games, a player is defined by a sequence of
observations, a collection of estimates, and a sequence of actions,
all of which may be highly complex. We now introduce a definition of a
player:

\begin{definition}
  \label{def:player}
  We call $(\cO, \fL_\varphi, \U)$ a player in the environment
  $(\O^u,\cF^u, \dbP^u)$ with observations $\cO$, learning algorithm
  $\fL_\varphi$, and with behavior $\U$, where
  \begin{equation}
    \begin{aligned}
      &\cO &&: \Omega^u\times \sA\times\dbN \to \sE,
      \\&
      \fL_\varphi &&: \sE \times \sM_\varphi \times \sM_\U \to \sM_\varphi,
      \\&
      \U &&: \sE \times \sM_\varphi \times \sM_\U \to \sM_\U.
    \end{aligned}
  \end{equation}
\end{definition}
From the perspective of mathematics, the question is what natural
conditions can be imposed on this collection of functions defining a
player. We begin with the consistency condition for the observations:
\begin{definition}
  \label{def:consistent-observations}
  We say a player has consistent observations, if
  \begin{equation*}
    \begin{aligned}
      &\cO(\o^u, a_\cdot, n)\in\sE
      \ \  \text{is a subsequence of}\ \
      \cO(\o^u, \tilde a_\cdot, n+1)\in\sE
      \\&\text{if}\ \ a_\cdot\in\sA
      \ \ \text{is a subsequence of}\ \
      \tilde a_\cdot\in \sA, \ \ \text{for all}\ \  n\in \dbN, \o^u \in \O^u
    \end{aligned}
  \end{equation*}  
\end{definition}
We note that $\cO$ sets the connection between the environment and the
player, and is not available for the player to evaluate.

Next, we define a recurrence condition for a player’s behavior. To do
so, we first introduce the following definition:
\begin{definition}
  We call
  \begin{equation*}
    {}^n\U: \O^u\times \dbN\to \sM_\U
  \end{equation*}
  the planned behavior of the player at age $n$, where
  \begin{equation*}
    \begin{aligned}
      &{}^n\U &&:= \U({}^n\cO, {}^n\fL_\varphi, {}^{n-1}\U) \in \sM_\U,
      \\&
      {}^n\fL_\varphi &&:=
      \fL_\varphi({}^n\cO, {}^{n-1}\fL_\varphi, {}^{n-1}\U) \in \sM_\varphi,
      \\&
      {}^n\cO &&:= \cO(\o^u, {}^{n-1} I, n) \in \sE,
      \\&
      {}^n I &&:= ({}^1\U(\o^u,\cdot), \dots, {}^n\U(\o^u,\cdot)) \in \sA.
    \end{aligned}
  \end{equation*}
\end{definition}

These functions
${}^n(\U, \fL_\varphi, \cO, I) := ({}^n\U, {}^n\fL_\varphi, {}^n\cO,
{}^nI)$ with domain $\O^u\times\dbN$ are determined in the order
${}^{n-1}I \to {}^n\cO \to {}^n \fL_\varphi \to {}^n\U$. Behaviors in
$\sM_\U$ depend on $\O^u$, and evaluating at $\o^u$ yields a sampled
(or observed) action. That is, for ${}^k\U \in \sM_\U$, we denote
${}^k\U(\o^u,\cdot)$ as the action taken, suppressing the rest of the
unspecified domain. Let us point out that a player might have the
capacity to generate randomness independently of the surrounding
environment. For now, we do not explicitly track potentially
independent probability spaces, such as those a player might use to
sample randomness, but instead include them within the general
environment. Along these lines, only the component of $\o^u\in\O^u$
that is relevant to the random variable under consideration is taken
into account.

We are now ready to introduce an intrinsic concept for the behavior of
player. Let us equip the space $\sM_\U$ with a generic metric $d$ and
define:
\begin{definition}
  \label{def:recurrence}
  \sloppy We say ${}^*\U\in \sM_\U$ is a $(r,\d)$-recurrent behavior
  for a player, if
  \begin{equation*}
    \dbP^u\Big(\liminf_{n\to\infty} d({}^*\U, {}^n\U) > r \Big) \leq \d.
  \end{equation*}
  Also, we say ${}^*\U$ is almost surely a recurrent behavior of the
  player if $r=\d=0$.
\end{definition}
In words, we classify behaviors that may occur infinitely often as the
player ages.

Finally, let us turn to the more intricate task of imposing conditions
on estimates. Motivated by the brain’s predictive nature, we introduce
a notion of a player that estimates future abstract representations of
observations. This lies at the core of behavior, and, roughly
speaking, decision-making is about forming preferences over future
observations. If we formalize observations as states, rewards, and
actions, then it becomes crucial to understand the future states,
rewards, and actions of other players. Then, we designate preferred
future observations as those with higher total rewards. Given that
observations are high-dimensional and complex, a player may need to
simplify the task. For example, with visual observations, instead of
predicting future pixels directly, one typically first forms useful
embeddings to facilitate prediction. In case of actions, a player
might estimate only intensions or goals of an oponnent. A similar
reduction applies to rewards, rather than predicting future rewards
directly, one may aim to learn the expected future reward. A more
sophisticated agent might aim to learn a distribution of rewards.

To add structure, we introduce objects and relations formed from
observations as the first layer of estimates. Let
$\dbN_\obj\subset \dbN$ be an index set for different objects. For
each $j\in\dbN_\obj$, let $E_\obj^j$ denote the space of states for
object $j$. Finally, let $E_\rel$ be a space representing the set of
relations. Then, let
\begin{equation*}
  \begin{aligned}
    &\sM_\obj := \big\{\sE \to \Pi_jE_\obj^j \big\},\ \ 
    \sM_\con := \big\{ \Pi_jE_\obj^j \to 2^{\dbN_\obj} \big\},
    \\[1.1ex]&\text{and}\quad
    \sM_\rel := \big\{ \Pi_jE_\obj^j \times 2^{\dbN_\obj} \to E_\rel \big\}.
  \end{aligned}
\end{equation*}
Correspondingly, we have the learning algorithms
\begin{equation*}
  \fL_{\{\obj,\con,\rel\}}:
  \sE\times \sM_{\varphi}\times \sM_{\U}\to \sM_{\{\obj,\con,\rel\}}
\end{equation*}
Then, ${}^n\fL_\obj\in\sM_\obj$ is a mapping from observations to
states of identified objects. The mapping ${}^n\fL_\con\in\sM_\con$
creates connections between those objects. Finally,
${}^n\fL_\rel\in\sM_\rel$ assigns relations to connections, which may
take the form of discrete tags or numerical values. All of these
learning algorithms may depend on past observations, estimates
(denoted generically by $\varphi$), and the behavior mapping.

Now, the one-step prediction problem can be formulated as the task of
modeling the next objects and relations. To represent this, let us
introduce
\begin{equation*}
  \sM_\pre :=
  \big\{\O^u\times \Pi_jE_\obj^j\times 2^{\dbN_\obj}\times E_\rel \times \dbA
  \to \Pi_jE_\obj^j\times 2^{\dbN_\obj}\times E_\rel\big\}
\end{equation*}
along with the learning algorithm $\fL_\pre$ and its realization
${}^n\fL_\pre\in\sM_\pre$ at age $n$.
\begin{definition}
  \label{def:one-step-predictive}
  We say
  $(\varphi_\obj, \varphi_\con, \varphi_\rel, \varphi_\pre) \in
  (\sM_\obj, \sM_\con, \sM_\rel, \sM_\pre)$ is one-step
  $\e$-predictive under some metric $d$ on
  $\cP(E_\obj\times 2^{\dbN_\obj}\times E_\rel)$, if
  \begin{equation*}
    \begin{aligned}
      &\hspace{0em}
      \liminf_{n\to\infty}
      d\big(\ \mathrm{Law}({}^{n+1}E | {}^n\cO, {}^n I),\
      \mathrm{Law}(
      \varphi_\pre(\o^u, {}^n E, {}^n \U(\o^u,\cdot))| {}^n\cO, {}^n I
      )\ \big)\leq \e,
      \\[1.2ex]&\text{where}\ \ 
      {}^n E := \big(\varphi_\obj({}^n\cO),\ 
      \varphi_\con(\varphi_\obj({}^n\cO)),\ 
      \varphi_\rel(\varphi_\obj({}^n\cO),
      \varphi_\con(\varphi_\obj({}^n\cO)))\big)
    \end{aligned}
  \end{equation*}
  $\dbP^u$-almost surely.
\end{definition}
We note that $\varphi_{\pre}$ is typically a random variable whose law
models the distribution of ${}^{n+1}E$. Let us also point out a
potentially confusing notation. Given $({}^n\cO, {}^nI)$, the realized
action ${}^n\U(\o^u,\cdot)$ is determined, whereas
$\varphi_\pre(\o^u,\cdot)$ typically includes an independent component
to model a distribution.

Let us discuss some examples to motivate the definitions above. Some
related references will be provided in Section \ref{sec:Control}.

\nob We can simplify by considering the overall state of the
environment as an observation, treated as a single object without
further decomposition and connections. And, $\varphi_\rel$ may assign
values to those states. In this case, $\varphi_\pre$ models the
one-step transition probabilities of the whole state, viewed as a
random variable, and the next value associated with it.

\nob Encoding the external environment is typically necessary. For
humans, for example, the eyes encode visual observations from the
environment. In the context of machine learning, a suitable neural
network architecture carries out the encoding. Various objectives may
be assigned to encoders, and $\varphi_\pre$ represents the predictive
one, which might be the next embedding, a reward component of the
observations, or another useful aspect of the observations for a given
task.

\nob Instead of encoding the environment as a single object, one may
construct many objects, each with its associated states. Then, for
example, $\varphi_\con$ may form pairs, and $\varphi_\rel$ may assign
attention values.

\nob Similarly, in the context of large language models (LLMs),
objects can correspond to token embeddings. Here, $\varphi_\con$ forms
pairs, connecting each future token to all previous ones (forming a
triangular matrix), and $\varphi_\rel$ uses keys and queries to assign
weights to these connections. This constitutes only a small component
of such models.

\nob In the context of chess, objects can be defined as pieces
together with their positions on the board. A player may then assign a
large number of connections between such objects through a highly
dynamic $\varphi_\con$, for example between pieces that protect each
other or among larger groups representing the overall
arrangement. Then, $\varphi_\rel$ may assign values to each
connection, and $\varphi_\pre$ may model the behavior of the opponent.

Let us emphasize that the aim of this work is not to design new
learning algorithms for estimating future observations. Our objective
is to reframe common game settings from the perspective of the player
and to introduce a concept of equilibrium via conditions on
$(\cO, \cL_\varphi, \U)$. Furthermore, we emphasize that designing a
player can be far more involved than setting up the game itself. We
argue that understanding complex interactions between players cannot
be reduced to the external setting of the game under consideration.

When there are multiple players, we keep track of them using the index
$i\in\dbN_0:=\{1,2,\dots\}$. For any symbol $\psi$ introduced in the
definitions above, we set $\vec\psi := \prod_{i\in\dbN_0}\psi^i$. In
this case, $\cE^i$, the space of observables for player $i$, usually
includes the actions or states of the other players. They interact and
influence one another, and their collective recurrent behaviors
${}^*\vec\U$ form a basis for an equilibrium.

Lastly, we introduce notions of cooperation and competition defined by
the players’ behavior rather than by the structure of the game
itself. The goal is to separate these concepts from the external
setting and use them to clarify our objectives when designing
players. In essence, we ask whether the players we model are meant to
exhibit stable, convergent behaviors, or to continually generate
diversity through dynamic interaction.
\begin{definition}
  We say players $(\vec\cO, \vec\fL_\varphi, \vec\U)$ in the
  environment $(\O^u, \cF^u, \dbP^u)$ are eventually cooperating at
  ${}^*\vec\U\in \vec\sM_\U$, if
  \begin{equation*}
    \dbP^u\Big(\limsup_{n\to\infty} \sup_{i\in\dbN_0}
    d({}^*\U^i, {}^n\U^i) > 0\Big) = 0.
  \end{equation*}
  We say that players are indefinitely competing, if for any
  $\vec\U\in\vec\sM_\U$,
  \begin{equation*}
    \dbP^u\Big(\limsup_{n\to\infty} \sup_{i\in\dbN_0}
    d(\U^i, {}^n\U^i) > 0\Big) = 1.
  \end{equation*}
\end{definition}

As an illustration, consider a simple setting of algorithmic price
competition between two firms. Although the environment is designed to
be competitive, if each firm’s pricing algorithm learns that the
competitor’s prices tend to move in a positively correlated manner,
both algorithms can quickly converge to stable, elevated price
levels. In this case, we regard the firms as cooperating, even though
the underlying game remains competitive.

\section{Discrete Games}
\label{sec:Discrete-Games}

Let $\dbT = \dbN$ denote the time indices, and $\dbS_t$ be a
measurable state space for each $t\in\dbT$. Set
$\dbS:= \bigcup_{t\in\dbT}\dbS_t$. For arbitrary set $E$ with Borel
$\sigma$-algebra $\cB(E)$, let $\cP(E)$ denote the set of all
probability measures on $E$. We will always consider discrete indexing
to avoid discussions on regularities and measurability.

Take $\O := \prod_{t \in \dbT} \dbS_t$ as the canonical
space. Define $X : \dbT \times \Omega \rightarrow \dbS$ as the
canonical process: $X_t : \Omega\to \dbS_t$, and
$X_t(\omega) = \omega_t$ for each $\omega \in \dbX$ and $t \in
\dbT$. Let $\dbF^X$ denote the filtration generated by $X$.  We always
require any function defined on $\dbT\times \Omega$ to be Markovian,
similar to the canonical process, and denote their parameters as
$(t,x)$ where it is understood that $x\in\dbS_t$.

We use $i\in\dbN_0:= \dbN\setminus \{0\}$ as the index for
players. For any $i\in\dbN_0$, let $\dbA^{t,x;i}$ be the action space
of player $i$ at $(t,x)\in\dbT\times\dbS$. Introduce
\begin{equation*}
  \vec\dbA^{t,x} := \prod_{i\in\dbN_0}\dbA^{t,x;i},\qquad
  \vec\dbA := \bigcup_{(t,x)\in\dbT\times\dbS} \vec\dbA^{t,x},\qquad
  \dbA^i := \bigcup_{(t,x)\in\dbT\times\dbS} \dbA^{t,x;i}.
\end{equation*}
Let us also introduce the space of controls;
\begin{equation*}
  \cA^i := \big\{ \a:\dbT\times\Omega \to \dbA^i\ : \
  \a(t,x) \in \dbA^{t,x;i} \ \ \forall (t,x)\in\dbT\times\dbS_t \big\},
  \quad\forall i\in\dbN_0
\end{equation*}
and set $\vec\cA := \prod_{i\in\dbN_0}\cA^i$.

For the connection between players and the environment, let $\cE^i$
denote the space of observables for player $i$, and set $\sE^i$ as the
finite sequences of $\cE^i$. Similar to previous notations, set
$\vec\cE := \prod_{i\in\dbN_0}\cE^i$ and
$\vec\sE := \prod_{i\in\dbN_0}\sE^i$. We will state every estimate as
depending on $(t,x)$, and hence we will assume
$\dbT\times\O\subset\cE^i$. Also, set $\sA^i$ as finite sequences in
$\dbA^i$, and $\vec\sA := \prod_{i\in\dbN_0}\sA^i$.

As for the learning parameters, we will now begin to introduce
horizon, transitions between states, transition costs, state values,
potential behaviors of other players, optimal controls and
expectations of the players. Our choices are inherently limited as a
player might be arbitrarily complicated. Our aim here is to restate
the general setting for many of our games in the perspective of player
, and demonstrate the concept of uncertain equilibrium. For
simplicity, we will temporarily disregard the index $i$ and focus
solely on the perspective of a single player.

First, we let players have a \emph{horizon} $\hat{T}$. Players cannot
predict the future indefinitely with reasonable accuracy. In other
words, as the horizon of prediction increases, the distribution of the
state process contains progressively less useful information,
eventually rendering it useless. Thus, let
\begin{equation}
  \label{eqn:hatT}
  \sM_T := \Big\{ \hat T : \dbT\times\Omega \to \dbT \Big\}
\end{equation}
be the space of all such functions, where the corresponding learning
algorithm will take values in. Notice that we allowed the horizon to
depend on the state, since the player might be able to project further
in well-trained states. More importantly, rather than a fixed time,
one can consider a stopping time
$\hat T^{(t,x)}(s,y): (\dbT\times\Omega)^2\to \dbT$, where
$\hat T^{(t,x)}$ is a stopping time for the future estimated process
in (\ref{eqn:distributionP}).

Next, the player have an estimate of the transition probabilities;
\begin{equation}
  \label{eqn:hatp}
  \begin{aligned}
    &\hat p: \dbT\times \Omega \times \vec\dbA \times \dbS
    \to \dbR^+,\ \ \text{where}\\
    &\hat p(t,x, \vec a; \cdot)
    \ \text{ is a probability measure on }\ \dbS_{t+1},
    \\&
    \text{for all}\ t\in\dbT, x\in\dbS_t, \text{ and }
    \vec a \in \vec\dbA^{t,x}.
  \end{aligned}
\end{equation}
Similarly, introduce $\sM_p$ as the space of all such mappings in
\eqref{eqn:hatp}.

Given $\hat p$ as in \eqref{eqn:hatp}, an initial
$(t,x)\in(\dbT, \dbS_t)$, and $\vec\a\in\vec\cA$, player induces a
distribution $\dbP^{t,x,\vec\a}:= \dbP^{\hat p; t, x, \vec\a}$ for the
canonical process as usual; for all $t\leq s$ and
$(\tilde x,y)\in (\dbS_s,\dbS_{s+1})$, initial condition is
$\dbP^{t, x, \vec\a}(X_t = x) = 1$ and
\begin{equation}
  \label{eqn:distributionP}
  \dbP^{t, x, \vec\a}(X_{s+1} = y | X_s = \tilde x) =
  \hat p(s, \tilde x, \vec\a(s, \tilde x); y).% ,\ \ \text{and}\ \ 
\end{equation}
Note that relaxed controls further integrate over the distribution of
controls to define \eqref{eqn:distributionP}. We instead integrate the
value below.

It is crucial that players learn about other players' behavior. To
fully understand any complex game, we cannot overlook this
fact. Knowledge of opponents' strategies intrinsically alters the
observed events within the game. Even a player's value depends on it,
as different opponents might tend to employ varying
strategies. Consequently, the value associated with a strategy cannot
disregard the opponents' reactions. Thus, we assume that a player
learns potential controls of others based on their own control;
\begin{equation}
  \label{eqn:hatG}
  \hat \G^i : \dbT\times \cA^i \to \cP(\vec\cA)\quad
  \text{and set}\ \
  \hat \G_{t,\a}^i(d\vec\a)
  := \hat \G_t^i(\a;d\vec\a) := \hat \G^i(t,\a)(d\vec\a)
\end{equation}
Denote $\sM_\G$ as the space of all such mappings. We remark two
points for (\ref{eqn:hatG}):

\noindent (i) We assume that players model the others potential
controls depending on their own control. However, one might model that
this depends on the path of states of the players, or any other
observables are legitimate as long as the cost (\ref{eqn:MainValue})
is well-defined.

\noindent (ii) For competing players, a sophisticated player might
have an estimate on how their actions could be exploitable in order to
deceive an opponent, deviating their $\hat\G$ from their actual
planning. To not only compete with but also cooperate with other
players, they may need to generate reliable estimates of the actions
of others. In the two-player game discussed in Section
\ref{sec:TwoPlayerGame}, because the costs to the players depend on
each other's states, omitting this aspect from the player model won't
accurately capture the observed dynamics.

An important notion to introduce is the value of a player. As future
observations are ranked by some associated rewards, value function
captures a qualitative information about the future rewards for a
given strategy. Now, we introduce transition costs and state
values:\footnote{We use cost and value interchangeably. In the case of
  scalar objectives as in this work, distinction is more
  pronounced. However, for multi-objective frameworks, there is
  typically no binary choice, but rather a continuum of choices.}
\begin{equation}
  \label{eqn:hatFhatphi}
  \hat F:\hat \Omega\times \dbT\times\O\times\vec\dbA\to \dbR,\qquad
  \hat \phi:\hat \Omega\times \dbT\times\O\to \dbR,\ \
\end{equation}
and let $\sM_F,\sM_\phi$ denote the sets of mappings as in
\eqref{eqn:hatFhatphi}. An important difference is that the player
models these as random variables on some probability space
$(\hat \Omega,\hat \cF,\hat \dbP)$, which we are now explicitly
separating from the environment and view it as an independent
component of it. In particular cases, it might be useful to
characterize the measure space $(\hat\O,\hat\cF, \hat\dbP)$, however,
once can also fix a sufficiently large probability space and
concentrate on the random variables. We remark that state value
$\hat \phi$ induces an ordering on states, and reaching a particular
state by different intermediate paths, or different set of actions
might have varying costs, which is aimed to be captured by the
transition cost $\hat F$.

Now, given
$(\hat T, \hat p, \hat \G, \hat F, \hat \phi)\in
\sM_T\times\sM_p\times\sM_\G\times\sM_F\times\sM_\phi$, the value of
player becomes
\begin{equation}
  \label{eqn:MainValue}
  \begin{aligned}
    &J(t,x;\a) :=
    \int_{\vec\cA} J(t,x;\vec\a)\hat\G_t(\a;d\vec\a),
    \ \ \text{where denoting}\ \  \dbE^{t,x,\vec\a} := \dbE^{\dbP^{t,x,\vec\a}},\\
    &J(t,x;\vec\a) :=
    \dbE^{t, x , \vec\a}
    \Big[\hat \phi(t+\hat T, X_{t+\hat T})
    + \sum_{s=t}^{t+\hat T-1} \hat F(s, X_s , \vec\a(s,X_s))\Big],\quad
    % \hspace{-2em}
  \end{aligned}
\end{equation}
which is a random variable on $\hat\Omega$. Set $\sM_J^i$ as the space
of all such functions $(\hat\O\times\dbT\times\O\times\cA^i\to
\dbR)$. We point out that requiring a random variable instead of a
distribution allows us to refer to samples.

We remark again that a general abstract setting might be a
simplification, and there might be many layers of various estimations,
such as objects and their relations, before a player actually
constructs its value estimate. Indeed, it may be the case for every
other estimate too. Transition probabilities might be estimated from
simple frequency analysis, or could be modeled by large attention
architectures. In fact, a truly complex player, such as a human, would
not only adapt existing estimates and behaviors, but also develop new
types of estimates and behaviors over time. Just as a child starts
with limited capacities and gradually acquires a rich repertoire, such
dynamic structural growth might indeed be essential for modeling
higher-order intelligence. While this remains far beyond current work,
it represents a compelling direction for future research.

Let us recall the game of chess, which serves as an excellent example
to keep in mind throughout this work. In chess, \(\hat{p}\) yields
deterministic transitions. However, a player does not know what
actions the opponent will take within \(\{t, \dots, t + \hat T\}\),
and beyond that, it is unclear what the transition costs of actions or
the value of being in a particular state at \(t + \hat T\) might
be. These are all crucial components for a player to learn. Notably,
the heuristic values assigned to pieces are designed to guide players
in learning \(\hat{F}\) and \(\hat{\phi}\). While simplistic, these
heuristics serve as an initial guide. Moreover, as we have mentioned,
knowledge about the opponent can alter the values of strategies, which
is captured in (\ref{eqn:MainValue}) abstractly. Let us also emphasize
that the player's horizon may depend significantly on the current
state. Towards the endgame, for instance, a well-trained chess player
might be able to estimate many steps ahead, whereas this ability may
be considerably more limited during the middle stages of the game.

As the player faces the optimization problem (\ref{eqn:MainValue}), it
is not always feasible to solve for the optimal control. When
\(\hat{T} = 1\), the problem might be relatively simple, allowing for
straightforward searches for \(\epsilon\)-optimal actions. However,
for longer horizons, the space of potential controls becomes
excessively large, complicating the search for optimal solutions. To
formalize this, let us first define
\begin{equation*}
  \a =^{t,x;i} \tilde \a\quad \text{if}\quad
  \a(s,y) = \tilde\a(s,y)
  \quad\forall s\in \{t,\dots, t + \hat T^i(t,x) - 1\},\ y\in \dbS_s
\end{equation*}
Under this equivalency relation, we introduce the quotient space
\begin{equation*}
  \cA^{t,x;i} := \cA^i / =^{t,x;i}
\end{equation*}
And then, to incorporate the potential difficulty and uncertainty in
identifying the optimal control, we introduce the next policy
estimation;
\begin{equation}
  \label{eqn:hat-pi}
  \begin{aligned}
    &\hat \pi: \hat\O\times\dbT\times\O \to \cA^i
    \qquad\text{where},
    \\&\hat\pi(\hat\o,t,x)\in \cA^{t,x;i},\quad
    \forall (\hat\o,t,x)\in\hat\O\times\dbT\times\dbS_t
  \end{aligned}
\end{equation}
Here, at $(\hat\o,t,x)$, $\hat\pi$ approximates the potential optimal
controls for $J(\hat\o,t,x,\cdot)$, which will be dictated by the
equilibrium condition below. We remark that $\hat\O$ might have a
component for both value and policy, which we don't keep track
explicitly. For example, we might have a collection of events where
the value is determined, whereas $\hat\pi$ is still a random
variable. Set $\sM_\pi$ as the set of functions as in
\eqref{eqn:hat-pi}.

Now, even when optimal control can be solved exactly, uncertainty over
the value will naturally induce a probability distribution over
controls. First, introduce the space of behavior $\sM_\U$ as functions
of the form
\begin{equation}
  \label{eqn:hat-U}
  \hat\U:\hat\O\times\dbT\times \O \to \dbA^i
\end{equation}
Then, we introduce a behavior in a straightforward manner from the
current estimates as follow;
\begin{equation}
  \label{eqn:Thompson-U}
  \begin{aligned}
    &\U:\Pi_{\varphi}\sM_\varphi \to \sM_\U \quad \text{where, }
    \\& \U(\hat\pi)
    = ((\hat\o, t,x) \mapsto \hat\pi(\hat\o, t, x)(t,x))
    \quad 
  \end{aligned}
\end{equation}
Here, behavior is choosing the immediate action assigned by the
policy, and reflects the randomness induced by both the value and
policy. This behavior is typical for the learning or playing phase and
is, in essence, similar to Thompson sampling adapted to our
framework. During a competition phase, one might choose a different
$\U$ as deterministic, selecting the action corresponding to the mode
of them. In particular situations, such as performing surgery, it is
not only wrong but also unethical to forgo the most likely action and
instead select one at random. Moreover, a player may adjust its
behavior to occasionally select unlikely actions, exploring states
that are disadvantageous or even entirely unseen when facing a weak
opponent.

It is important to motivate the role of randomness in value, which
then induces a probability distribution over actions by
\eqref{eqn:Thompson-U}. Recall that in sufficiently complex settings,
such as chess, values are inherently unknown and must be learned
through significant effort. That is, the randomness of the value
models what is unknown to the player. One key role of randomness in
value is to allow players to explore systematically. If the player is
not satisfied with the current value estimates, it is natural to shift
the estimates for unexplored controls, or their outcomes, toward
higher values, in anticipation that they may achieve better results
than current estimates. This approach naturally leads the player to
search for controls yielding more satisfactory outcomes. We will
demonstrate a toy version for the two-player game in Section
\ref{sec:TwoPlayerGame}. This approach aligns with the common
intuition that a better understanding of values should lead to less
uncertain strategies.

Now that we have introduced the spaces of estimations and behavior,
let us turn to players as in Definition \ref{def:player}. Observations
may come from real-world experience, or, in the mean-field regime,
players can generate observations by assuming that every other player
is identical. For multiple players, we define the observations as
mappings of the form
\begin{equation}
  \begin{aligned}
    \label{eqn:observations}
    \cO^i : \Omega^u\times \vec \sA \times \dbN \to \sE^i
  \end{aligned}
\end{equation}
satisfying consistency as in Definition
\ref{def:consistent-observations}. We then set
$\vec\cO = (\cO^1,\cO^2,\dots)$.

Next, we formally acknowledge the existence of learning algorithms. We
say a collection of functions $\fL_{\varphi}^i$ for
$\varphi\in\{T,p,\G,F,\phi,\pi\}$ is the learning algorithm of player
$i$. Recall that $\sM_T,\sM_p, \sM_\G, \sM_F,\sM_\phi,\sM_\pi$
respectively denote the spaces of estimations as in \eqref{eqn:hatT},
\eqref{eqn:hatp}, \eqref{eqn:hatG}, \eqref{eqn:hatFhatphi}, and
\eqref{eqn:hat-pi} respectively.  Then, learning algorithms are in
general of the form
\begin{equation}
  \begin{aligned}
    \label{eqn:learningalgorithms}
    &\fL_\varphi^i : \sE^i\times\Pi_{\tilde\varphi} \sM_{\tilde\varphi}\times \sM_\U
    \to \sM_\varphi, \quad \forall \varphi\in\{T,p,\G,F,\phi,\pi\}.
  \end{aligned}
\end{equation}
Let us remark that, although it was not necessary, we introduced $\U$
explicitly given the other estimations. In its given form, it is a
value-based approach in reinforcement learning (see Section
\ref{sec:Control}). However, learning algorithms in
\eqref{eqn:learningalgorithms} are crucial, and we do not attempt to
simplify them. For example $p$ and $\G$ might be defined directly from
observations by keeping frequency statistics. Although this allows
them to be more trackable, they are limited to simple settings. Our
motivation in this work is to emphasize their inherent complexity
instead. They must be subject to evaluations of their respective
objectives, such as in Definition \ref{def:one-step-predictive}.

Lastly, to introduce the estimates and the planned behavior of
players, we denote the priors as
\begin{equation*}
{}^0 \vec I \in \vec\sA, \quad\text{and}\quad
  {}^0\vec\fL_\varphi \in \vec\sM_\varphi, \quad
  \forall \varphi\in\{T,p,\G,F,\phi,\pi\}.
\end{equation*}
This sets ${}^0\vec\U$ as in \eqref{eqn:Thompson-U}. Then,
\begin{equation}
  \label{eqn:discrete-planned-behavior}
  \begin{aligned}
    &{}^n\U^i &&:= \U({}^n\fL_\varphi^i) \in \sM_\U,
    \quad \forall i\in\dbN_0
    \\[0.7ex]&
    {}^n\fL_\varphi^i &&:=
    \fL_\varphi({}^n\cO^i,\ {}^{n-1}\fL_\varphi^i,\ {}^{n-1}\U^i) \in \sM_\varphi,
    \quad \forall \varphi\in\{T,p,\G,F,\phi,\pi\}, i\in\dbN_0
    \\[0.7ex]&
    {}^n\vec\cO &&:= \cO(\o^u, {}^{n-1} I, n) \in \vec\sE,
    \\[0.7ex]&
    {}^n\vec I &&:= ({}^0\vec I,
    {}^1\vec\U(\hat\o, t, x), \dots, {}^n\vec\U(\hat\o, t, x)) \in \vec\sA.
  \end{aligned}
\end{equation}
Also, ${}^n(\fL_T, \fL_p, \fL_\G, \fL_F, \fL_\phi)^i$ defines
${}^n J^i(t,x;\a)$ as in \eqref{eqn:MainValue}, and we may use the
notation ${}^n\hat\varphi^i := {}^n\fL_\varphi^i$ if convenient.

\begin{definition}[Uncertain Equilibrium of Discrete Games]
  \label{def:Uncertain-Equilibria}
  We say that players $(\vec\cO, \vec\fL_\varphi, \vec\U)$ admit
  ${}^*\vec\U\in\vec\sM_\U$ as an $(\e, r, \d)$-uncertain equilibrium
  under metrics $d^i$ on $\sM_\U^i$, if for any prior
  ${}^0 \vec I\in \vec\sA$,
  \begin{itemize}
  \item[(i)]
    \begin{equation*}
      \hspace{-2em}
      \limsup_{n\to\infty}\sup_{i\in\dbN_0}\int_{\hat\O}\Big(\
      \sup_{\tilde\a\in\cA^i} {}^n J^i(\hat\o,t,x,\tilde\a) -
      {}^nJ^i(\cdot,{}^n\fL_\pi^i)(\hat\o,t,x)\ \Big)
      \hat\dbP(d\hat\o) \leq \e,
    \end{equation*}
  \item[(ii)]
    \begin{equation*}
      \dbP^u\Big(
        \liminf_{n\to\infty} \sup_{i\in\dbN_0}
        d^i({}^*\U , {}^n\U^i) > r\Big)
      \leq \delta
    \end{equation*}
  \end{itemize}
\end{definition}
Note that condition (iii) aligns with the recurrence defined in
Definition \ref{def:recurrence}. We further impose condition (ii) on
estimations to obtain a more favorable notion of equilibrium. Later in
this section, we will introduce an additional learning parameter and
discuss how to incorporate it into this definition.

Let us revisit the example of chess. Consider a well-trained chess
player, and suppose the game is nearing its end. At this late stage,
there are often configurations in which subsequent moves are
certain. That is, a particular action has an induced probability of
one, and remains unchanged as the player continues to
learn. Similarly, at the opening of the game, the player might have a
distribution over different openings. Although we will not observe the
same opening in each game, for a well-trained player this distribution
may evolve only over long time scales. On the other hand, there may be
many configurations where learning continues indefinitely. We remark
that $d$ is an arbitrary metric on functions of the form
$\{\hat\O^i\times\dbT\times \O \to \dbA^i\}$, which can be designed to
reflect these considerations.

Notice that we require players to approach a particular behavior
independent of their previous history. Such independence implicitly
requires that players explore sufficiently diverse behaviors to be
able to reach this equilibrium. Furthermore, since players may be
exploitable, they often change their behaviors. However, an
equilibrium is one that recurs infinitely often.

To briefly elaborate on how players might solve their own optimization
problems, they may do so by revisiting past observations with evolving
estimations. As the player accumulates observations (that is, as $n$
increases) and recalls past observations, the exploration of potential
scenarios under the current estimations ${}^n\hat\varphi^i$ aims to
capture the term
$\sup_{\a \in \mathcal{A}^i} {}^nJ^i(\hat\o, t, x, \a)$. During this
revaluation and exploration process, new strategies may be discovered,
or an updated assessment of value might lead to changes in
\( {}^n\hat\pi^i \). Condition (ii) in the definition of uncertain
equilibrium implies that players have explored potential strategies
and are capable of generating the best ones under various scenarios of
\( \hat\O \) as learning continues.

On the other hand, the values of the player are driven externally.
Not only is there no universal notion of what holds higher value, but
values often involve multiple, conflicting objectives. These are
shaped by needs, interactions, and self-evaluations, as reflected in
the remarkable diversity of values across individuals and societies.
In settings that we design, scalar values are again externally
assigned to observations, and the value function is a representation
of these assignments.

The choice of $\liminf$ instead of $\limsup$ is important. As
mentioned in the Introduction and in Section
\ref{sec:defining-players}, we interpret convergence of behavior as
cooperation between players. By using $\liminf$, we require that a
particular behavior remains favorable and is used infinitely often,
though not necessarily always.

Let us briefly explain the role of $(\e, r, \d)$, which we take to be
uniform over players for simplicity. Firstly, $\e$ characterizes how
effectively players can solve their respective optimization
problem. Next, $r$ measures how closely the distribution of strategies
approaches the equilibrium infinitely often. Lastly, $\d$ reflects the
likelihood that an equilibrium will be observed. Recall that $\dbP^u$
is associated with the universe in which the players exist; this may
depend on the real underlying dynamics, and random choices generated
by each player.

We remark that the planned behavior ${}^n\U^i(\cdot, t,x)$ in
\eqref{eqn:discrete-planned-behavior} is a function of the form
$\dbN\times \O^u\to \dbA^{t,x;i}$, thus a discrete-time stochastic
process taking values in the space of actions available at
$(t,x)$. From this perspective, an uncertain equilibrium can be viewed
as a recurrent point of this process. The primary interest lies in
understanding the evolution of this process under a specific design of
player.

We now clarify the similarities to and differences from the concept of
correlated equilibrium. To do so, we examine the optimality condition
in Definition \ref{def:Uncertain-Equilibria} of uncertain equilibrium
(simplifying notation by omitting n n) as follows:
\begin{equation*}
  \begin{aligned}
    \int_{\hat\O} J^i(\hat\o,\hat\pi^i(\hat\o))
    \hat\dbP(d\hat\o)
    = \int_{\hat\O}
    \int_{\vec\cA}J^i(\hat\o,\vec\a) \hat\G^i(\hat\pi^i(\hat\o); d\vec\a)
    \hat\dbP(d\hat\o)
  \end{aligned}
\end{equation*}
and since $\hat\G^i\in \cP(\vec\cA)$ (see \eqref{eqn:hatG}),
\begin{equation}
  \label{eqn:induced-distribution-over-vecA}
  \int_{\hat\O}\hat\G^i(\hat\pi^i(\hat\o);d\vec\a)
  \hat\dbP(d\hat\o) \in \cP(\vec\cA)
\end{equation}
To connect with the concept of correlated equilibrium, consider any
$\rho\in\cP(\vec\cA)$. By disintegrating $\rho$ with respect to its
$i$-th component as
$\rho(d\vec\a) = \rho^{-i}(d\vec\a|\a^i)\rho^i(d\a^i)$, we identify
that $\rho^{-i}$ corresponds to $\hat \G^i$ and $\rho^i$ corresponds
to $\hat\dbP(\hat\o)$.\footnote{One can extend $\rho^{-i}$ to the full
  $\vec\cA$ by the Dirac distribution on $\a^i$ for the $i$-th
  marginal, which is also the case for $\hat\G^i$.} Roughly speaking,
the equilibrium conditions can then be expressed (using simplified
notations) as follows:
\begin{equation*}
  \begin{aligned}
    &\text{Nash-type Equilibrium:}\quad
    &&\int_{\cA^i}\int_{\vec\cA} \sup_{\tilde\a^i}J^i(\tilde\a^i,\vec\a^{-i})
    \rho^{-i}(d\vec\a|\a^i)\rho^i(d\a^i)
    \\&\text{Correlated Equilibrium:}\quad
    &&\int_{\cA^i}\sup_{\tilde\a^i}\int_{\vec\cA} J^i(\tilde\a^i,\vec\a^{-i})
    \rho^{-i}(d\vec\a|\a^i)\rho^i(d\a^i)
    \\&\text{Uncertain Equilibrium:}\quad
    &&\int_{\hat\O}\sup_{\tilde\a^i}\int_{\vec\cA} J^i(\hat\o,\tilde\a^i,\vec\a^{-i})
    \hat\G^i(\tilde\a^i;d\vec\a)\hat\dbP(d\hat\o)
    \\&\text{Coarse Correlated Equilibrium:}\quad
    &&\sup_{\tilde\a^i}\int_{\cA^i}\int_{\vec\cA} J^i(\tilde\a^i,\vec\a^{-i})
    \rho^{-i}(d\vec\a|\a^i)\rho^i(d\a^i)  
  \end{aligned}
\end{equation*}
The supremum over $\tilde\a^i$ has different dependence in each
equilibrium concept. In the Nash-type equilibrium\footnote{We refer to
  this as a Nash-type equilibrium due to its structure; however, as it
  is potentially impossible to satisfy, such a concept does not, to
  our knowledge, appear in the literature.}, it depends on
$\vec\a^{-i}$; in the correlated equilibrium, it depends on $\a^i$; in
the uncertain equilibrium, it depends on $\hat\o$; and in the coarse
correlated equilibrium, it is independent of the controls. We replace
$\hat\pi$ with $\sup_{\tilde\a^i}$ to reflect the optimality
condition, and $\hat\dbP$ plays a role analogous to $\rho^i$. However,
the key difference is that the correlated equilibrium, similar to the
Nash equilibrium, considers deviations that do not affect other
players, whereas in uncertain equilibrium, changing one's strategy
influences others via learned estimations.

Recall that if regret, defined analogously to the correlated
equilibrium, is sublinear, then the time-averaged empirical
distribution of actions converges to a correlated
equilibrium. Considering the estimated distributions over strategies
in (\ref{eqn:induced-distribution-over-vecA}), one might expect all
players to eventually induce the same distribution in symmetric
situations. However, in general, there is no reason to expect that a
single distribution characterizes every player’s
considerations. Moreover, a similar criticism applies to regret
definitions, since changing prior actions would typically influence
the future strategies of others.

The concept of Nash equilibrium focuses solely on controls, according
to their associated scalar values, inherently excluding the intrinsic
structure of a player. For example, in situations where a central
planner announces policies for individual agents, such as
environmental regulations, traffic management, public health
initiatives, with the knowledge that every individual will act
according to their own assessment of value, the Nash equilibrium is
the appropriate framework. The central planner needs to model agents'
individual values to construct stable policies. In such cases, it is
not meaningful to model each and every player in detail through their
learning algorithms. We also remark that the Nash equilibrium requires
players to have exact knowledge of the strategies of others. In the
pure equilibrium sense, this instability is significant enough that an
equilibrium may fail to exist even in the simplest games. To address
this, one typically adopts relaxed controls, which is essential,
though not motivated by player design.

Let us highlight the importance of incorporating additional learnable
parameters into our framework. This could include encoding raw
observations, establishing communication protocols, and many other
spaces of estimations to design more sophisticated player. One such
simple yet interesting parameter is a player's expectation regarding
the best achievable outcome, which defines a notion of regret for the
player and alters the characteristics of exploration, as demonstrated
in simplified form in Section \ref{sec:TwoPlayerGame}. Consider
functions of the form
\begin{equation}
  \label{eqn:best-expectation}
  \hat B: \dbT\times\O \to \dbR
\end{equation}
and all the related definitions similar to other learning parameters.
Then, we can introduce;
\begin{equation*}
  {}^n\k^i(t,x) := \hat\dbP\bigg(
  {}^nJ^i(\cdot,{}^n\hat\pi^i)(\hat\o,t,x) >
  {}^n\hat B^i(t,x) \bigg)
\end{equation*}
To relate this quantity to familiar concepts with which we all can
relate, we say that at the state $(t,x)\in\dbT\times\O$, the player
$i$ is currently desperate if ${}^n\k^i(t,x) = 0$, and euphoric if
${}^n\k^i(t,x) = 1$. If the player is desperate, as the learning
progresses, either ${}^n\hat B^i$ will decrease, leading the player in
some sense to accept the situation, or ${}^nJ^i$ will assign higher
values to underexplored strategies, encouraging the player to explore
them. One can further describe the player's current situation using
verbal subcategories like
\begin{equation*}
  \small
  \text{Desperate} - \text{Discouraged} - \text{Doubtful} -
  \text{Determined} -
  \text{Confident} - \text{Optimistic} - \text{Euphoric}
\end{equation*}
which can be interpreted as partitions of $\k$-values. Beyond
providing a richer characterization of a player, this notion can be
incorporated into Definition \ref{def:Uncertain-Equilibria} of
equilibrium by requiring
\begin{itemize}
\item[(ii')]
  \begin{equation*}
    \limsup_{n\to\infty} \sup_{i\in\dbN_0} {}^n\k^i(t,x) > \k ,
  \end{equation*}
\end{itemize}
and denoting it as $(\k,\e,r,\d)$-uncertain equilibrium. In other
words, we now search for player designs that further achieve
confidence.

In addition to the design of the player, the choice of equilibrium is
also diverse. For example, to recover the concept of Nash equilibrium,
we can assume
\begin{itemize}
\item[(ii'')] for all $i\in\dbN_0$ and $n\in\dbN$ large enough,
  \begin{equation*}
    {}^n\hat\G^i(t,x) =
    \text{Law}({}^n\hat\pi^1(\cdot, t,x))\times 
    \text{Law}({}^n\hat\pi^2(\cdot, t,x))\times\cdots
  \end{equation*}
\end{itemize}
where we extend $\hat\G$ to depend naturally on $x$, while removing
dependence on $\a$. That is, players' estimates yield the strategies
of others, effectively meaning that players are able to observe each
other's future strategies. If every estimate is predetermined as part
of the setting and no randomness is involved, then, together with
condition (ii) in Definition \ref{def:Uncertain-Equilibria}, we
recover that the players’ policies $\hat\pi^i$ form a Nash
equilibrium.

Next, time-consistency, or Dynamic Programming Principle, can be
naturally introduced for any learning parameter and can likewise be
required at equilibrium. The most important and familiar one is for
the value function: We say that the estimate
$(\hat T, \hat p, \hat \G, \hat F, \hat \phi, \hat \pi)$ yields a time
consistent value almost surely, if for any $(t,x)\in\dbT\times\dbS_t$
and $0\leq T_0\leq \hat T(t,x)$, it holds $d\hat\dbP$-a.s. that
\begin{equation}
  \label{eqn:time-consistent-value}
  \int_{\cA^i} J(T_0;\hat\o,t,x,\a) \hat\pi(\hat\o,t,x)(d\a) =
  \int_{\cA^i} J(\hat\o,t,x,\a) \hat\pi(\hat\o,t,x)(d\a)
  % \quad d\hat\dbP-a.s. \hspace{-3em}
\end{equation}
where $J(T_0;\hat\o,t,x,\a)$ is defined exactly as in
(\ref{eqn:MainValue}), but $\hat T$ is replaced by $T_0$. Notice that
when $T_0=t$ and $\hat\pi$ yields the optimal control for each
$\hat\o\in\hat\O$, (\ref{eqn:time-consistent-value}) becomes
\begin{equation*}
  \begin{aligned}
    &\hat\phi(t,x) =
    \sup_\a J(t,x,\a)
    \\&= \sup_\a
    \int_{\vec\cA}
    \dbE^{t, x , \vec\a}
    \Big[\hat \phi(t+\hat T, X_{t+\hat T})
    + \sum_{s=t}^{t+\hat T-1} \hat F(s, X_s , \vec\a(s,X_s))\Big]
    \hat\G_t(\a;d\vec\a)
  \end{aligned}
\end{equation*}
which closely resembles the standard time-consistency, or Dynamic
Programming Principle, for the standard value function.

To illustrate how time consistency can be required of an estimation,
we focus on $\hat\pi$; similar conditions can be formulated for
$\hat T$, $\hat p$, and $\hat\G$ as well.\footnote{In our case,
  $\hat p$ is a single-step transition probability and is therefore
  automatically consistent.} The idea is straightforward: the
distribution induced by $\hat\pi(\cdot, t,x)$ given a future state
$(T_0,x_{T_0})$, should be the same as if the policy were formed at
the state $(T_0, x_{T_0})$. Formally, we say that the estimate
$(\hat T, \hat p, \hat \G, \hat F, \hat \phi, \hat \pi)$ yields a
time-consistent distribution over controls almost surely at
$(t,x)\in \dbT\times\dbS_t$ if, for any
$t\leq T_0\leq t + \hat T(t,x)-1$, and $x_{T_0}\in\dbS_{T_0}$,
$\hat \pi(\hat\o, t,x)$ induces the same distribution as
$\hat \pi(\hat\o, T_0, x_{T_0})$ on the space
$\cA^{\hat T(t,x);T_0,x_{T_0},i}$. Here, the quotient space is defined
similarly, with the relation terminating at $t + \hat T(t,x) -1$
instead of $t + \hat T(T_0,x_{T_0}) -1$.

\subsection{Two player game example}
\label{sec:TwoPlayerGame}

In this section, we present a simple, repeatedly played two-player
example. We demonstrate that even with a fixed one-step horizon,
players can exhibit non-stationary dynamics. In this setting, both
players learn transition costs $\hat{F}$ and the actions of their
opponents $\hat\G$, all within a fixed horizon $T=1$. Our central
argument is that formulating controls as an equilibrium does not
adequately capture the dynamic strategies continually employed by the
players. To address this shortcoming, we construct learning algorithms
that capture these dynamic strategies. This example illustrates why a
more general framework is necessary for effectively modeling games,
and it also heuristically highlights how the concept of equilibrium is
inherently shaped by the learning process and by the opponents
themselves.

Consider fixed state and action spaces given by $\dbS = \{0,1\}$ and
$\dbA=[0,1]$. Players' actions determine their transition
probabilities at each step, and they can only observe each other's
state. The first player loses $\$1$ if the second player appears in
state $1$ but gains by increasing their own transition probability to
state $1$. The second player loses $\$1$ if they are not in the same
state. \footnote{We mention their gains and losses in dollar amounts,
  to relate easily to a potential game we might play in real life.}

Let us note that, with a one-step horizon, there exists a unique Nash
equilibrium that the first player is unwilling to play. Of course, by
virtue of Folk's theorem, any feasible outcome can be sustained in an
infinitely repeated game. However, this result explicitly relies on
the assumption that players are certain about their opponents' future
actions over an indefinite horizon. This strong assumption allows for
almost any feasible value to be supported as an equilibrium, leaving
the question of which outcome will be observed without a clear answer.
Moreover, since players do not announce their strategies, searching
for a Nash equilibrium with a larger horizon does not necessarily
model this game either. Such a search represents our external attempts
to formalize the players’ incentives. Instead, we emphasize once again
that the core element in games is the learning algorithms employed by
the players. These algorithms naturally govern the (random) evolution
of probability distributions over actions, which in turn is sufficient
to understand the evolution of the game.

Now, let us formally state the game. Consider fixed state and action
spaces given by $\dbS = \{0,1\}$ and $\dbA = [0,1]$. Suppose the
players are not learning the horizon or transition probabilities, that
is, $\fL_T^i,\fL_p^i$ are constants yielding $\hat T^i=1$ and
$p^i(t,\vec x,\vec a, 1) = a^i$. Initially, to specify the rules of
the game and enable comparison with the Nash equilibrium, we assume
that the players are not learning the state value or transition costs;
\begin{equation*}
  \begin{aligned}
    &\phi^1(t+1, X_{t+1}^1,X_{t+1}^2) = -\ind{X_{t+1}^2=1},\hspace{2em}
    F^1(t,X_t^1,X_t^2, a^1, a^2) = c a^1,\\
    &\phi^2(t+1, X_{t+1}^1,X_{t+1}^2) = -\ind{X_{t+1}^1 \neq X_{t+1}^2},
    \hspace{0.5em} F^2 = 0
  \end{aligned}
\end{equation*}
As mentioned, the first player wants the second player to move state
$0$, and the second player wants to be in the same state as the
first. However, since $c>0$, the first player gains by increasing the
oods of moving to state $1$. Now, costs of each player are
\begin{equation}
  \label{eqn:2Player-nonrandomJ1J2}
  \begin{aligned}
    &J^1(t,\vec x, \vec a) = ca^1 - \dbP^{t,\vec x, \vec a}(X_{t+1}^2 = 1),\ \ 
    J^2(t,\vec x, \vec a) = -\dbP^{t,\vec x, \vec a}(X_{t+1}^1 \neq X_{t+1}^2),
    % \ \ \text{ where, }
    \\[1.5ex]&
    \dbP^{t,\vec x, \vec a}(X_{t+1}^2 = 1) = a^2,\hspace{2em}
    \dbP^{t,\vec x, \vec a}(X_{t+1}^1 \neq X_{t+1}^2) = a^1 + a^2(1-2a^1)
  \end{aligned}
\end{equation}
From now on, since the one-step game does not depend on the current
time or state, we will drop them from notations. Let us also note that
players make their decisions simultaneously. One could instead
formulate the game as turn-based, but we aim to keep it as symmetric
as possible, while excluding only the cost structure.

Note that when the horizon is fixed at $1$, there exists a unique Nash
equilibrium given by $\vec{a} = (1,1)$. Although such an equilibrium
exists, it is not necessarily useful for characterizing the potential
behaviors of the players. Once the first player fixes the action of
the second player, they lose awareness of the latter’s underlying
intentions.

Let us recap how the game is played from the perspective of the first
player. First, we determine a probability $a^1$ of transitioning to
state $1$, and receive a payoff $ca^1$. Then, we lose
\$$1$ if the other player ends up in state
$1$. From the perspective of the second player, objective is simply to
follow the other player. We observe the past states of the other and
attempt to end up in the same state, losing \$$1$ otherwise. Due to
the simplicity of the game, behaviors that are expected to recur over
time can be generated. Starting with the second player, whose cost
structure is simpler:
\begin{itemize}
\item[$\bullet$] Determine an acceptable level of noise in observing
  the other player's state, based on expectations. If recent
  observations consistently yield a particular outcome ($0$ or $1$)
  within the acceptable noise level, take the corresponding
  action. Otherwise, begin exploring other actions, with rationale to
  penalize the noise.
\end{itemize}
For the first player, the cost structure is slightly more intricate;
\begin{itemize}
\item[$\bullet$] If the second player consistently appears in state
  $0$, explore larger actions to reduce the cost (due to
  $ca^1$). Continue increasing it until the second player begins to
  appear in state $1$ frequently enough to offset these gains.

\item[$\bullet$] If the second player appears at $1$, which is costly,
  switch to actions that are not used recently. Continue exploring
  until the second player reappears in state $0$ regularly enough to
  keep the realized cost consistent with expectations.
\end{itemize}

In the next section, we construct a learning algorithm and numerically
explore how the corresponding behaviors evolve. We remark that a
straightforward $Q$-learning algorithm can be used to model the
players. However, $Q$-learning models only expected rewards for each
action, it therefore converges to fixed actions and lacks the
underlying dynamics we aim to demonstrate. This underscores the same
point: it is crucial to incorporate the design of the player to
understand games.

\subsubsection{Details of the Learning Algorithm}
We now introduce the relevant parts of the framework specific for this
problem. In general, each player models a transition cost $\hat F$ and
an estimate $\hat\G$ on the other player's actions, relying on their
observations held in the memory. Also, each player has an expectation
$\hat B$ as in \eqref{eqn:best-expectation}, taken as a constant for
simplicity.

Recall that observations of players are the realized states. That is,
$\cE^1 = \cE^2 = \dbS$, and observations in \eqref{eqn:observations}
are depending on the realizations of the states. Here, $\O^u$ and
$\dbP^u$ are determined by the random number generators that
determines the transitions of states for the players at each round. We
then set the observations $\cO^1,\cO^2$ in equation
(\ref{eqn:observations}) in an obvious manner. In the simulations,
each player keeps a memory of a certain length, recording the realized
states and costs.

Now, let $\big\{\cN_\G^{i,k}\big\}_{k=1}^K$ be the $i$'th player
simple feed-forward networks where $\cN_\G^{i,k} : \dbA \to \dbA$. We
then assing $\fL_\G^i: \sE^i \to (\dbA \mapsto \cP(\dbA))$ in
\eqref{eqn:learningalgorithms} as the empirical distribution formed by
$\{\cN_\G^{i,k}\}_k$'s. That is,
\begin{equation*}
  {}^n\hat\G^i := \fL_\G^i({}^n\cO^i)
  := \frac{1}{K}\sum_{k=1}^K \d_{\cN_\G^{i,k}}
\end{equation*}
Note that the parameters of networks are depending on the
observations, which is not explicit in notations as we view
$\fL_\G^i$ yielding the network with such parameters. To train
these networks, after each step, players draw a batch of memories
using the multinomial distribution with higher weights assigned to
recent observations. Then, networks are getting trained to reduce the
difference between estimated action and observed state.

Similarly, we denote $\big\{\cN_F^{i,\ell}\big\}_{\ell=1}^K$, where
$\cN_F^{i,\ell}:[0,1]\to \dbR$. We then set
$\hat F: \hat\O \times \dbA \to \dbR $ in \eqref{eqn:hatFhatphi} as
\begin{equation*}
  \hat F^1(\ell, \hat\o', a^1) = ca^1 + \cN_F^{1,\ell}(\hat\o')(a^1),\quad
  \hat F^2(\ell, \hat\o', a^2) = \cN_F^{2,\ell}(\hat\o')(a^2)
\end{equation*}
We identify $(\ell,\hat\o')\in\hat\O = \{1,\dots,K\} \times \hat\O'$
where $\hat\dbP$ assigns the first marginal as uniform distribution
over $\{1,\dots,K\}$.\footnote{As in the case of action networks, this
  is only for simplicity. One might assign and evolve weights
  corresponding to networks, and capture more vibrant dynamics if the
  game is more sophisticated. For example, one might keep a subset of
  networks as trusted ones (high weights), and let other networks
  explore more wildly (low weights).} Each network $\cN_F^{i,\ell}$ is
further random by the virtue of dropout layers. Keeping the networks
always in the training mode, one generates a random function with
positive dropout probabilities, and $\hat\O'$ abstracts this. Here,
\begin{equation*}
  \fL_F^i :\sE^i \to \big((\hat\O, \dbA) \mapsto \dbR\big)
\end{equation*}
yielding ${}^n\hat F^i$ and $\phi^i$'s are taken as constant.

There are two objectives cost networks are training for: (i) there is
an expected cost coming from the predictions of action networks, which
is (\ref{eqn:2Player-nonrandomJ1J2}) integrated as in
(\ref{eqn:MainValue}). If action networks are not perfect, expected
cost will not match the observed expected cost. Cost networks are
training to close this gap, by relying on costs in the memory. (ii):
players have expectations over what is best possible as introduced in
(\ref{eqn:best-expectation}), which we took as constant for
simplicity. Cost networks are also trained such that the players do
not get desperate. For example, if the first player ends up with
networks $\cN_\G^{1,k}$'s taking values close to $1$, independent of
the action, they start to play the Nash equilibrium $(1,1)$. That is,
player $1$ gets desperate, and then adjusts the random component of
cost to increase values of other actions towards $\hat B^1$ to start
exploring them.

We note that (\ref{eqn:MainValue}) becomes
\begin{equation*}
  \begin{aligned}
    & J^1(\ell, \hat\o', a^1)
    = \frac{1}{K}\sum_{k=1}^K
    \cN_\G^{1,k}(a^1) + c a^1 + \cN_F^{1,\ell}(\hat\o')(a^1)
    ,\ \text{ and}
    \\& J^2(\ell, \hat\o', a^2)
    = \frac{1}{K}\sum_{k=1}^K
    \cN_\G^{2,k}(a^2) + a^2(1-2\cN_\G^{2,k}(a^2)) + \cN_F^{2,\ell}(\hat\o')(a^2)
  \end{aligned}
\end{equation*}
To draw an action from \eqref{eqn:Thompson-U}, players uniformly
choose one cost network $\ell \in \{1,\dots,K\}$, and observe one
realization $\hat\o'$ coming from dropout layers. Then,
$\U^i = \hat\pi^i$ simply minimizes $J^i$ over $a^i$ yielding
$\e$-optimal action, and plays it.

Before discussing the simulation results, let us mention that each
parameter of networks of course plays a significant role, and we
coarsely tuned them by hand to obtain simulations matching with
expectations. Many of such parameters are taken as constant. Changing
to different constants might of course yield poor results. On the
other hand, generalizing them will improve the sophistication of
agents strategies. Besides the network parameters, there are more
structural parameters too. For example, what players are expecting as
the best possible cost also changes the characteristics of
actions. Especially if the first player is expecting much better than
what is realistically possible, exploration gets out of hand. For an
another example, if the memory is very long and not forgetting, then
the first player starts to get advantage over the second player, as
the second player becomes fixed after a while and estimates that the
other player will still frequently move to state $0$. The point we are
aiming to convey here is that the learning algorithms of players are
crucial to characterize what is going to be realized in reality.

\begin{figure*}[h!]
    \centering
    \begin{subfigure}[b]{0.47\textwidth}
        \centering
        \includegraphics[width=\textwidth]{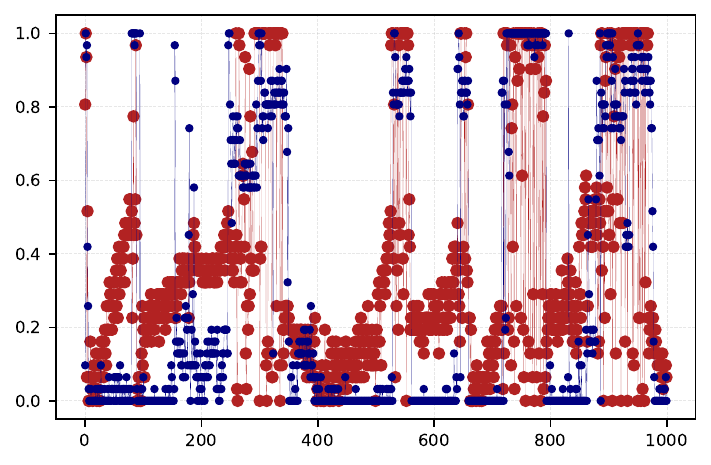}
        \caption{$(c, B^1, B^2) =
          \big(\tfrac{3}{10}, \tfrac{1}{10}, -\tfrac{1}{5}\big)$}
        \label{fig:plot1}
    \end{subfigure}
    \begin{subfigure}[b]{0.47\textwidth}
        \centering
        \includegraphics[width=\textwidth]{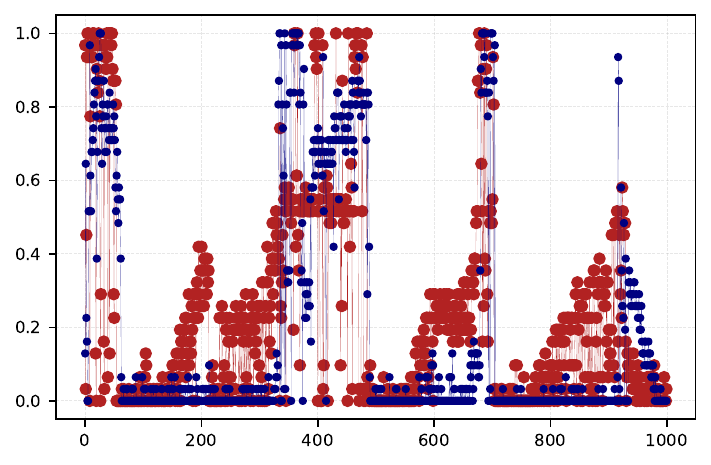}
        \caption{$(c, B^1, B^2) =
          \big(\tfrac{1}{20}, -\tfrac{1}{10}, -\tfrac{1}{5}\big)$}
        \label{fig:plot2}
    \end{subfigure}
    
    \begin{subfigure}[b]{0.47\textwidth}
        \centering
        \includegraphics[width=\textwidth]{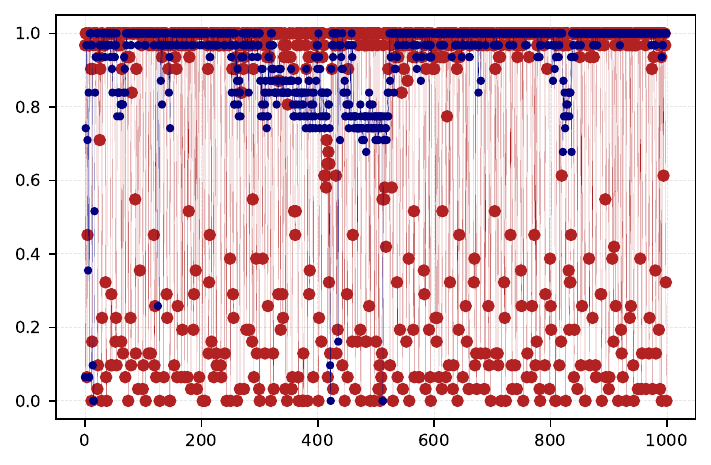}
        \caption{$\big(c, B^1, B^2) = (1, 1, -\tfrac{1}{5}\big)$}
        \label{fig:plot3}
    \end{subfigure}
    \begin{subfigure}[b]{0.47\textwidth}
        \centering
        \includegraphics[width=\textwidth]{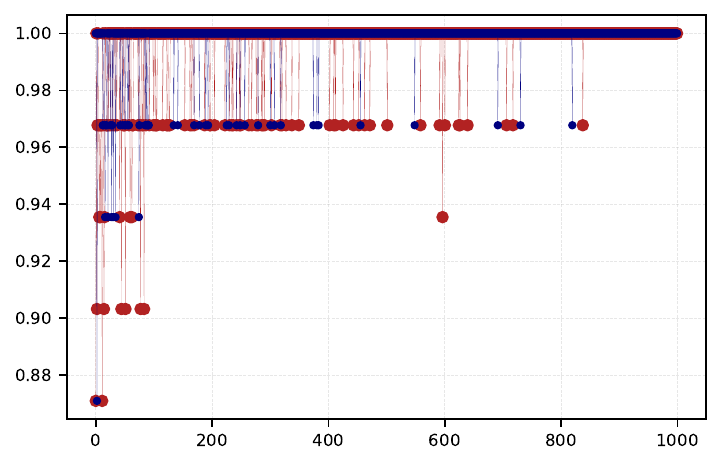}
        \caption{$\big(c, B^1, B^2) = (1, 0, -\tfrac{1}{5}\big)$}
        \label{fig:plot4}
    \end{subfigure}
    \caption{Actions of Player $1$ (red) and Player $2$ (blue) over
      $1000$ games, demonstrating the impact of varying the incentive
      parameter $c$ and expectation $B^1$ for Player $1$, while Player
      $2$'s parameters remain constant. Each subplot shows how
      different incentives and expectations influence Player $1$'s
      strategy and interaction dynamics in this toy problem.}
    \label{fig:2PlayerGame-all-actions}
\end{figure*}

Now, let us annotate the simulation results. In Figure
\ref{fig:2PlayerGame-all-actions}, the actions taken by Player $1$
(red) and Player $2$ (blue) over $1000$ games are plotted, with both
players starting from arbitrary initial estimations. While keeping
Player $2$'s parameters constant, four plots illustrate variations in
$c$, the incentive parameter for Player $1$, and $B^1$, as defined in
(\ref{eqn:best-expectation}), which represents the expectation of
Player $1$. Thin lines in the plots indicate the jumps between
actions.

In subplot (a), Player 1 has a somewhat large incentive ($c = 3/10$)
to take larger actions, aiming for a reward of $B^1 = 1/10$. Thus,
playing close to $(0,0)$ does not suffice, and Player 1 searches for
higher rewards, leading to frequent changes between different
phases. Notice that as soon as $(a^1, a^2)$ approaches $(1,1)$, Player
1 begins to explore and pushes the game back towards $(0,0)$. In
subplot (b), the incentive is much smaller ($c = 1/20$) and the
expectation of Player 1 is decreased to $B^1 = -1/10$. Consequently,
Player 1 stays close to $(0,0)$ for longer, before starting to think
that Player 2 will always choose action $0$. In subplots (c) and (d),
the incentive for Player 1 is really high $(c=1)$, making deviations
from $(1,1)$ unnecessary. Specifically, in subplot (d), Player 1 is
satisfied with a reward of $0$, maintaining $(1,1)$ almost
always. Conversely, in subplot (c) where Player 1 expects to get an
unrealistic reward of $1$, which requires Player $1$ to play $1$ while
Player $2$ plays $0$, exploration by Player 1 leads to worse results
for both.

We refer to the supplementary online repository \cite{TwoPlayerGame}
for the animation showing the cost networks, action networks, and
other observables in each case. We point out that these estimations
are not converging networks, instead dynamic and yielding repeating
patterns of behaviors.

We conclude this section by emphasizing once more that games are
inherently complex and that observed behaviors require a detailed
representation of players. In our working paper \cite{BIM2025}, we
explore firm collusion in an online price competition setting. In
Calvano et al. \cite{CCDP2020}, the authors model each firm using
$Q$-learning. Each firm maintains a single $Q$-function estimate,
requiring an intensive training period to populate the $Q$ table, an
approach that is infeasible in online learning situations and offers
little intuition about why collusion emerges. In \cite{BIM2025}, we
design firms as outlined in this section. Beyond observing rapid
convergence, we aim to explain how collusion, or competition, is
driven by firms’ design choices, potentially offering policymakers
better guidance for regulating price competition.

\section{Stated Mean-Field Games}
\label{sec:stated-mean-field-games}

In this section, we will introduce a mean-field type version of the
framework. It is important to note that learning parameters are
defined for each player individually. Therefore, embedding a
mean-field game requires adjusting the learning parameters of a
representative agent to model infinitely many similar players. In
particular, to align with a similar structure in the literature, we
assume that only $\hat{\G}$ will be learned, while other parameters
$(\hat T, \hat p, \hat F, \hat\phi) = (T, p, F, \phi)$ are modeled by
the representative player as known (and not learned). We will thus
refer to this case as the stated mean-field game.

Let the state space be $\dbS_t$,
$\dbS := \bigcup_{t\in\dbT}\dbS_t$, and $\dbA$ be the common
action space. Set the canonical space
$\O := \prod_{t\in\dbT} \dbS_t$ and introduce the set of
controls as
\begin{equation*}
  \cA := \big\{ \a: \dbT\times \O \times \cP(\O) \to \dbA
  \ :\ \a(t,x,\mu) \in\dbA,\
  \forall (t,x,\mu) \in \dbT \times \dbS_t \times \cP(\dbS_t)\big\}
\end{equation*}
As before, we require any function on $\dbT\times\O\times\cP(\O)$ to
be Markovian. Transition probabilities are given as
\begin{equation*}
  \begin{aligned}
    &p(t,x,\mu,a; y):
    \dbT\times \O \times \cP(\O) \times \dbA 
    \times \dbS\to \dbR^+,\quad \text{where}
    \\&p(t,x,\mu,a; \cdot) \text{ is a probability measure on $\dbS_t$,
      for all $t\in \dbT, x\in \dbS_t, \mu\in \cP(\dbS_t)$}
  \end{aligned}
\end{equation*}

Next, along the idea that the representative agent is insignificant in
the population, we assume $\hat \G$ in (\ref{eqn:hatG}) is constant as
$\hat\G: \dbT \to \cP(\cP(\dbS_t\times \cA))$. Here,
$\cP(\dbS_t\times \cA)$ corresponds to $\vec\cA$ in
(\ref{eqn:hatG}). In the case of countable players, indexing was
keeping track of the connection between the state and the control of
individual players. Here, state variable keeps track of distribution
of controls used by the population.

Now, given a particular estimation of the population
$\Xi_t \in \cP(\dbS_t\times \cA)$ by the representative player at time
$t$, introduce $\Xi_s \in \cP(\dbS_s \times \cA)$ recursively as
\begin{equation}
  \label{eqn:MFG-populationflow}
  \Xi_{s+1}(dy, d\a) =
  \int_{\dbS_t}
  p(s,x,\mu_s^\Xi,\a(s,x, \mu_s^\Xi); dy) d\Xi_s(x,d\a),
  \ \forall t\leq s, 
\end{equation}
where $\mu_s^\Xi := \Xi_s(\cdot,\cA)$. Note that $\mu^\Xi$ corresponds
to (\ref{eqn:distributionP}) for the population. If the second
marginal of $\Xi$ is a Dirac measure $\delta_\a$ independent of the
state, we call it homogeneous, as it models every individual player
using a single control $\a$. Otherwise, we call it heterogeneous. In
the homogeneous case, we do not need to keep track of the flow of the
distribution of controls. Moreover, in the heterogeneous case, one can
represent the flow of the population $\mu^\Xi$ using a single relaxed
control instead of a distribution of controls. See \cite{IZ2024} for
the details.

Next, introduce the flow of the distribution for the representative
player;
\begin{equation}
  \label{eqn:MFG-playerflow}
  \dbP^{t,\Xi;x,\a}(X_{s+1} = dy| X_s = \tilde x)
  = p(s,\tilde x,\mu_s^\Xi,\a(s,\tilde x, \mu_s^\Xi); dy)
  \ \ \forall t\leq s,\ \ 
\end{equation}
with initial condition $\dbP^{t,\Xi;x,\a}(X_t = x) = 1$ where $X$ is
the canonical process. Notice that the player is observing the
distribution of the population $\mu^\Xi$, given the initial data
$\Xi\in \cP(\dbS_t\times \cA)$.

Recall that we assume the cost is known and not learned. Moreover,
while defining (\ref{eqn:MainValue}), we started from the initial
state $x$, and here we similarly start from the current distribution
$\mu \in \cP(\dbS_t)$. We will restrict the learning algorithm to
yield $\hat{\G}$ with its marginal on $\dbS_t$ as a Dirac measure at
$\mu$. Then, similar to (\ref{eqn:MainValue}), we assume the cost
structure is given by
\begin{equation}
  \label{eqn:MFG-J}
  \begin{aligned}
    &J(t,\mu ; x,\a) :=
    \int_{\cP(\dbS_t\times \cA)}
    J(t,\Xi; x,\a) d\hat \G_t(\Xi),\ \ \text{where}\ \
    \hat\G_t((\mu,\cP(\cA))) = 1, \text{ and} 
    \\&J(t,\Xi; x,\a) :=
    \dbE^{t,\Xi;x,\a} \Big[\phi(X_{t+T},\mu_{t+T}^\Xi) +
    \sum_{s=t}^{t+T-1} F(s,X_s, \mu_s^\Xi, \a(s,X_s,\mu_s^\Xi)) \Big],
    % \quad \dbE^{\cdot} := \dbE^{\dbP^{\cdot}}
  \end{aligned}
\end{equation}
Set $\sM_J$ as the space of all such functions
$(\dbT\times \cP(\O)\times \O \times \cA \to \dbR)$. Let us note that,
we are mainly interested in the static $\{0,\dots, T\}$ problem for
simplicity. One can dynamically set $\hat T = T-t$ (and repeats after
$T$) by the learning algorithm to create a dynamic version. Or one
might evolve the game indefinitely, keeping the $\hat T$ fixed if the
structure allows it.

Given the cost, we now need to estimate the optimal controls by the
learning parameter
\begin{equation*}
  \hat \pi: \dbT\times \cP(\O) \times \O \to \cA
\end{equation*}
There is no randomness in the value, and assuming that the value is
sufficiently representative, we don't further impose randomness in the
policy. Thus, given that representative player is able to solve for
the optimal control, $\hat\pi$ becomes deterministic, taking values on
the set of optimal controls. Moreover, behavior is simply
$\U(\hat\pi) = ((t,\mu;x) \mapsto \hat\pi(t,\mu;x))$. Due to the
time-consistency, optimal policy determined at an initial condition
stays optimal, but we will rely only on $\hat\pi$ to generate
observations rather than the behavior.

Let us briefly recap the mean-field framework. We assume that the
representative player starts with an initial guess of the population
distribution over states and controls, determines the corresponding
optimal control, and, relying on the assumption that everyone else is
exactly the same, generates further observations using the chosen
learning algorithm. Equivalently, one can say that there are
infinitely many such players actually playing the game and observing
the distribution of players, however, our framing is more consistent
with applications. For example, let us consider a portfolio
liquidation problem that someone faces in a financial market. Instead
of solving an optimization problem without acknowledging that there
are other players also facing a similar problem, as a first order
approach without actually having information about other players, the
player can model there is a distribution of others facing exactly the
same problem. 

Finally, we are ready to introduce the observations and the learning
algorithm. Let $ \cE = \cP(\cP(\dbS \times \cA)) $ be the space of
observables, and let $\sE$ denote the space of finite sequences of
$\cE$. Recall that the learning algorithm is a mapping
$\fL_\varphi: \sE \to \sM_\varphi$, where the player's estimation
at age $n$ is $ {}^n\fL_\varphi := \fL_\varphi({}^n\cO)$. Here,
$\cO: \dbN \to \sE$ represents the increasing sequence of
observations.  To provide intuition, we will construct a simple but
explicit learning algorithm for $\fL_\G$:

Suppose that the current distribution of the population at time $0$ is
$\mu\in\cP(\dbS_0)$ and is fixed as given. We, as the representative
player, start with an initial guess
${}^0\cO = {}^0\Xi = \d_{(\mu, \d_{{}^0\a})}$ for some ${}^0\a$. That
is, our initial observation is $\delta_{{}^0\Xi}$. Then, we determine
the population flow $\mu^{{}^0\Xi}$ using
(\ref{eqn:MFG-populationflow}), our flow $\dbP^{t,{}^0\Xi;x,\tilde\a}$
using (\ref{eqn:MFG-playerflow}), and solve the optimization problem
to find an optimal control ${}^1\a$.

Now, following the fixed point idea, we learned that if the population
is using ${}^0\a$, it is optimal to use ${}^1\a$. Since every player
is equivalent, we may deduce that the population will use ${}^1\a$
with some probability $c$, and use ${}^0\a$ otherwise. That is, we set
the learning algorithm as
\begin{equation*}
  \fL_\G(\delta_{(\mu,\delta_{{}^0\a})}) = c \delta_{(\mu, \delta_{{}^1\a})}
  + (1-c) \delta_{(\mu, \delta_{{}^0\a})} = {}^1\cO
\end{equation*}
Notice that, for simplicity, we are assuming a homogeneous
population. That is, everyone is assumed to be using a single
control. Once can, for example, formulate it as a portion of the
population will use ${}^1\a$. Next, we can repeat the same procedure
to find another optimal control under the guess
$\hat\G = \fL_\G(\delta_{(\mu,\delta_{{}^0\a})})$, denoted as
${}^2\a$, and so on. In general, our naive learning algorithm depends
only on the last observation, defined as
\begin{equation}
  \label{eqn:learning-algorithm-simple}
  \fL_\G\big({}^n\cO\big) :=
  c\ \delta_{(\mu, \delta_{{}^{n+1}\a})}
  + (1-c)\ \ {}^n\cO,\ \ 0\leq c\leq 1
\end{equation}
where we took ${}^n\cO \in \cP(\cP(\dbS_0 \times \cA))$ as a single
observation rather than a sequence to simplify notation, and
${}^{n+1}\a$ is an optimal control under ${}^n\cO$.

Let us remark on the similarity between the fictitious play-type
algorithms introduced in Cardaliaguet-Hadikhanloo \cite{CH2017}. In
fictitious play, one also starts with an initial guess $\d_{{}^0\a}$
and finds the optimal ${}^1\a$. A crucial difference, however, is that
fictitious play considers the weighted average of $\mu^{{}^0\Xi}$ and
$\mu^{{}^1\Xi}$ to find the next optimal control $\a^2$. That is, the
cost structure becomes
\begin{equation*}
  J^{\text{fictitious}}(t,\mu ; x,\a) :=
  J\bigg(t,\int_{\cP(\dbS_t\times \cA)}\Xi d\hat \G(\Xi); x,\a\bigg)
\end{equation*}
for $J$ as in (\ref{eqn:MFG-J}), with the $\hat \G$ induced by the
same $\fL_\G$ but solving a different optimization. We leave the
question of whether these approaches are equivalent for potential
games to future research, which is a key assumption in \cite{CH2017}
for the convergence result.

Lastly, we rephrase the definition of uncertain equilibrium with
notations as in Section \ref{sec:defining-players}, and we explicitly
compute the equilibrium under this basic learning algorithm in the
next section.
\begin{definition}[Uncertain Equilibria of stated Mean-Field Games]
  We say that a player $(\cO, \fL_\G, \fL_\pi, \U)$ admit
  ${}^*\U\in\sM_\U$ as an $(\e,\d)$-uncertain equilibrium under the
  metric $d$ on $\sM_\U$ at
  $(t,x,\mu)\in \dbT\times\dbS_t\times\cP(\dbS_t)$, if for any prior
  ${}^0\a\in \cA$,
  \begin{itemize}
  \item[(i)]
    \begin{equation*}
      \limsup_{n\to\infty} \big(\sup_{\tilde\a\in\cA} {}^nJ(t,\mu;x,\tilde\a)
      - {}^nJ(\cdot, {}^n\hat\pi)(t,\mu;x)\big) \leq \e
    \end{equation*}
  \item[(ii)]
    \begin{equation*}
      \liminf_{n\to\infty} d({}^*\U, {}^n\U) \leq \d
    \end{equation*}
  \end{itemize}
\end{definition}

\subsection{One step stated mean-field game examples}

We now present two examples in which we can explicitly compute and
contrast the relaxed equilibrium and uncertain equilibrium under the
learning algorithm described in
(\ref{eqn:learning-algorithm-simple}). In the first example, while
there is no standard Nash equilibrium, a relaxed equilibrium does
exist. Conversely, in the second example, due to the cost function
being discontinuous, there is no relaxed equilibrium; however, the
uncertain equilibrium remains unchanged.
\begin{example}
  \label{exa:mean-field-1}
  Set $\dbS = \{0,1\}$, $\dbT = \{0,1\}$, the action space
  $\dbA = [0,1]$, and the transition probability
  \begin{equation*}
    p(0, x, a, \mu; 1) = a,\qquad  p(0, x, a, \mu; 0) = 1-a
  \end{equation*}
  Furthermore, introduce the cost as
  \begin{equation*}
    \begin{aligned}
      &J(\Xi; \a) := \dbE^{\dbP^{\Xi,\a}}
      \Big[\phi(X_1, \mu_1^\Xi) + F(\a)\Big]
      \\&\text{where}\quad
      \phi(x,\mu) :=  2 |\mu(1)|^2 - 4\ind{x = 1}\mu(1)
      ,\ \ \text{and}\ \ F(a) = (|a|^2 + a)
    \end{aligned}
  \end{equation*}
  Then, there exists no standard Nash equilibrium and a unique relaxed
  equilibrium $\frac{1}{2}(\delta_0 + \delta_1)$.
  The learning algorithm described in
  \eqref{eqn:learning-algorithm-simple} oscillates around
  $\frac{1}{2}(\delta_{\delta_0} +
  \delta_{\delta_1})\in\cP(\cP(\dbA))$, and induces an action
  distribution $\delta_0$ or $\delta_1$ infinitely often.
\end{example}
\begin{proof}
  First, let us argue that there exists no standard Nash
  equilibrium. Main idea is, if the population distribution is
  symmetric $\mu_1^\Xi(1)=\mu_1^\Xi(0)=1/2$, then the optimal actions
  are $\{0,1\}$. Whenever $\mu_1^\Xi(1)>1/2$, optimal action becomes
  $0$ and otherwise $1$. That is, every player tries to stay away from
  the majority and there cannot be a deterministic fixed point.

  As for the standard Nash equilibrium population is homogeneous,
  (\ref{eqn:MFG-populationflow}) simplifies to
  \begin{equation*}
    \mu^a(1) := \mu_1^\Xi(1) = a
  \end{equation*}
  whenever the population is taking the action $a\in \dbA$,
  independent of the initial distribution. For the representative
  player, we reserve $\tilde a = \a(0,x)$ and compute the cost;
  \begin{equation*}
    \begin{aligned}
      J(a,\tilde a) := J(\Xi;\a) &=
      2|\mu^a(1)|^2 - 4\mu^a(1)\dbP^{\Xi,\a}(X_1 = 1) + |\tilde a|^2 + \tilde a
      \\&=
      2a^2 - 4a\tilde a + |\tilde a|^2 + \tilde a
    \end{aligned}
  \end{equation*}
  since it is quadratic in $\tilde a$, maximum occurs only if
  $\tilde a\in\{0,1\}$. Thus, noting that
  \begin{equation*}
    J(0,\tilde a) = |\tilde a|^2 + \tilde a,\quad
    J(1,\tilde a) = 2 + |\tilde a|^2 - 3 \tilde a
  \end{equation*}
  there exists no standard Nash equilibrium.

  Now, to compute the relaxed equilibrium, we know from \cite{IZ2024}
  that it is equivalent to consider the heterogeneous case. Thus, as
  the initial distribution is irrelevant, let
  $\Xi_0 \in \cP(\cA)=\cP(\dbA)$. Since there is only a single time
  step, the distribution of controls doesn't evolve either and
  (\ref{eqn:MFG-populationflow}) becomes
  \begin{equation*}
    \begin{aligned}
      &\Xi(1,da) := \Xi_1(1,d\a) = a \Xi_0(da),
      % \qquad
      \\&\mu^\Xi(1) := \mu_1^\Xi(1) = \Xi(1,\dbA) = \int_{[0,1]} a\Xi_0(da)
    \end{aligned}
  \end{equation*}
  Then,
  \begin{equation*}
    J(\Xi_0; \tilde a) := J(\Xi; \a) =
    2|\mu^\Xi(1)|^2 - 4|\mu^\Xi(1)|\tilde a + |\tilde a|^2 + \tilde a
  \end{equation*}
  and in this case, again from \cite{IZ2024}, equilibrium means that
  every action in the support of $\Xi_0$ is optimal. It is easy to
  check that if $\mu^\Xi(1) \neq 1/2$, then the optimal action is
  either $0$ or $1$ and there cannot be any equilibrium. If
  $\mu^\Xi(1) = 1/2$, then both $0$ and $1$ are optimal. Thus,
  $\Xi_0 = \tfrac{1}{2}(\delta_0 + \delta_1)$ corresponds to the
  relaxed equilibrium, since any action in the support is optimal.

  Lastly, let us discuss the convergence of
  (\ref{eqn:learning-algorithm-simple}). Consider any
  $\G = \sum_i c_i\delta_{\delta_{a^i}}$ which is an element of
  $\cP(\cP(\dbA))$ if $\sum_ic_i=1$ representing any homogeneous
  estimate for the action of the population. Then, under appropriate
  notational simplifications of this example, (\ref{eqn:MFG-J})
  becomes
  \begin{equation*}
    \begin{aligned}
      J(\tilde a) &= \int_{\cP(\dbA)} J(\Xi,\tilde a) d\G(\Xi)
      = \sum_i c_i J(\delta_{a_i},\tilde a)
      \\&= 2\sum_ic_i |a_i|^2 - 4\tilde a\sum_ic_i a_i + |\tilde a|^2 + \tilde a
    \end{aligned}
  \end{equation*}
  which is exactly as before a quadratic polynomial in $\tilde a$,
  hence optimal value occurs at either $0$ or $1$. Therefore, the
  algorithm (\ref{eqn:learning-algorithm-simple}) will quickly
  converge to a distribution having $\delta_0,\delta_1\in\cP(\dbA)$ in
  its support, and the contribution from the initial guess will
  diminish with the factor $(1-c)^n$. Moreover, as the optimal
  $\tilde a$ becomes $0$ or $1$ depending on the estimated average of
  the population similarly as before, the algorithm will oscillate
  around $\frac{1}{2}(\delta_{\delta_0} + \delta_{\delta_1})$. Note
  that, by adjusting the constant parameter $c$ in
  (\ref{eqn:learning-algorithm-simple}), one can achieve exact
  convergence too\footnote{Let us briefly take attention to the
    importance of the design of the learning algorithm, even for this
    simple setting. If the parameter $c$ diminishes very fast with
    $n$, then one can conclude either $\delta_0$ or $\delta_1$ is
    optimal depending on the initial condition.}. Here, since
  $\hat\pi$ is computed exactly as either $\delta_0$ or $\delta_1$, we
  also observe that the induced distribution oscillates in between
  them infinitely often.
\end{proof}

\begin{example}
  \label{exa:mean-field-2}
  Set $\dbS = [0,1]$, $\dbT = \{0,1\}$, the action space
  $\dbA = [0,1]$, and the transition probability
  \begin{equation*}
    p(0, x, a, \mu; dy) = \delta_{a}
  \end{equation*}
  Furthermore, introduce a discontinuous cost as
  \begin{equation*}
    J(\Xi;\a) := \dbE^{\dbP^{\Xi;\a}}\bigg[
    X_1^\a
    \ind{\bar\mu_1^\Xi \in \big[0,\tfrac{1}{2}\big]}
    - X_1^\a
    \ind{\bar\mu_1^\Xi \in \big(\tfrac{1}{2},1\big]}\bigg]
  \end{equation*}
  where $\bar\mu^\Xi := \int_{[0,1]} x\, d\mu^\Xi$.  Whereas no
  relaxed equilibrium exists, the learning algorithm described in
  \eqref{eqn:learning-algorithm-simple} again oscillates around
  $\frac{1}{2}(\delta_{\delta_0} +
  \delta_{\delta_1})\in\cP(\cP(\dbA))$, and induces an action
  distribution $\delta_0$ or $\delta_1$ infinitely often.
\end{example}
\begin{proof}
  The essence of this example is similar to that in Example
  \ref{exa:mean-field-1}. It is clear that there exists no relaxed
  equilibrium, as the optimal action is either $\delta_0$ or
  $\delta_1$ under any value of $\bar{\mu}_1^\Xi$, and neither
  constitutes an equilibrium.

  For the learning algorithm (\ref{eqn:learning-algorithm-simple}),
  although the cost function is computed differently than in Example
  \ref{exa:mean-field-1}, $\hat{\pi}$ behaves exactly the same,
  depending on the population average. Thus, there is no difference
  from Example \ref{exa:mean-field-1}.

\end{proof}

\section{Reinforcement Learning}
\label{sec:Control}

In this section, we review several core reinforcement-learning methods
using the terminology developed within this framework.  Our goal is
not to survey the extensive literature, but rather to emphasize the
complexity inherent in player design. In the following subsection, we
consider the framework of Section \ref{sec:Discrete-Games} in the
single-player setting to illustrate a learning method that does not
primarily rely on Bellman-type updates. In the following subsection,
we review multi-agent reinforcement learning methods.

Let us first review some main categories before detailing them
further.
\begin{itemize}
\item[(i)] Value-based design: The player $(\cO, \fL_\varphi, \U)$ is
  defined so that the learning algorithms take the form
  $\fL_\varphi:\sE\times\sM_\varphi\times\sM_\U \to\sM_\varphi$, and
  the behavior is given by $\U:\sM_\varphi\to\sM_\U$.  A simple
  example is $\sM_\varphi := \{\dbS\times\dbA \to \dbR\}$ and
  $\sM_\U := \{\dbS \to \dbA\}$, where $\U$ yields the maximizing
  argument over $\dbA$.
  
\item[(ii)] Policy-based design: The player $(\cO, \U)$ does not rely
  on an estimate but instead trains the behavior directly. Here, the
  behavior takes the form $\U:\sE\times\sM_\U\to\sM_\U$.
  
\item[(iii)] Actor-critic design: The player $(\cO,\fL_\varphi,\U)$ is
  defined so that the behavior takes the form
  $\U:\sE\times\sM_\varphi\times\sM_\U\to\sM_\U$. Here, estimations
  are trained similarly to value-based designs and are used to enhance
  the training of the policy-based behavior. The learning algorithms
  might also, and typically do, take the form
  $\fL_\varphi:\sE\times\sM_\varphi\times\sM_\U\to\sM_\varphi$ if they
  aim to learn the value estimate corresponding to the current
  behavior.
\end{itemize}

The simplest illustration is vanilla $Q$-learning.  The agent
maintains a single estimation space
\begin{equation*}
  \sM_Q := \big\{\,Q : \dbS\times\dbA\to\dbR\,\big\}  
\end{equation*}
and, after training, adjusts the behavior to act greedily:
\begin{equation*}
  \U : \sM_Q\to \sM_\U
  \qquad
  \U(Q) = \argmax_{a\in\dbA} Q(\cdot,a) \in \{\dbS \to \dbA\}
\end{equation*}
The learning algorithm,
$\fL_Q\colon\sE\times\sM_\varphi\times\sM_\U\to\sM_Q$, implemented,
for example, via temporal-difference updates, iteratively adjusts the
estimates to assign higher values to favorable outcomes while
preserving their time consistency. In this method, enforcing
consistency before the value estimates are well trained often leads to
instability. To mitigate this, one may introduce a slower-moving
“target” network to anchor the estimates while continuing to explore,
employ two value networks with decoupled or conservative targets to
reduce overestimation bias or apply additional regularization terms to
stabilize both the estimation and behavioral updates. See, for
example, \cite{Mnih2015,vieillard2020munchausen, vanhasselt2016deep,
  fujimoto2018addressing, haarnoja2018soft} for foundational and
modern off-policy implementations.

In on-policy learning, the estimates aim to capture the value of the
current behavior. This mitigates instabilities that arise from
enforcing consistency early in training, but at the expense of making
past observations less useful because the behavior continually
evolves. See, for example,
\cite{williams1992simple,sutton2000policy,mnih2016asynchronous}. For
training such estimates, which we do not elaborate on here, see also
\cite{schulman2017proximal,schulman2016high} for important
methodological considerations.

Distributional RL addresses the setting in which estimates represent
distributions rather than expectations, with estimates taking the form
$\sM_\varphi:\{\dbS\times \dbA \to \cP(\dbR)\}$. See, for example,
\cite{bellemare2017distributional, dabney2018distributional,
  dabney2018implicit, yang2019fully}. In the next subsection, we
present a toy example in which the estimates are random variables,
$\sM_\varphi: \{\hat \O\times \dbS \to \dbR\}$, similar to the concept
of ensemble methods such as \cite{osband2016deep}

Model-based RL incorporates further spaces of estimations for the
upcoming observations. As we briefly mentioned in Section
\ref{sec:defining-players}, such estimates may include future states,
rewards, and actions. Furthermore, one might first consider embeddings
of the observations to facilitate predicting future embeddings.  See,
for example, \cite{ha2018world, hafner2019learning, hafner2020dream}
for some modern implementations.

Once players have an estimate for future observations, they can plan
for the future in a time-inconsistent manner. Similar to the
consideration in Section \ref{sec:Discrete-Games}, suppose an
estimated policy takes values in the space of controls rather than
actions. In such cases, the plan devised for a potential future
observation may differ from what is actually planned when that
situation is realized. This is expected, as the player tailors its
efforts to the current situation. This is an important generalization,
which is typically absent in model-free approaches, and suggests that
time-consistency is an important but not fundamental property. We will
mention three approaches in this direction:

\nob Model Predictive Control and its modern applications in RL can be
seen as a direct example. See, for instance,
\cite{nagabandi2018neural, chua2018deep}.

\nob Monte Carlo Tree Search methods also plan into the future, using
policy solely to better evaluate the current possible actions; see,
for example, \cite{silver2016mastering, silver2018general,
  schrittwieser2020mastering}.

\nob Hierarchical methods include additional estimations that assign
goals to the behavior. This enables the generation of diverse
strategies and, in the case of model-free implementations, introduces
a form of time inconsistency. However, rather than discarding the
future plan, the player commits to the assigned goal. This provides
the flexibility to behave differently later, depending on the
previously assigned goal; see, for example, \cite{bacon2017option,
  nachum2018data, levy2019learning}.

Next, we briefly mention methods for learning representations from
observations. The main objective is to reduce the complexity of the
observations while retaining a representation that is sufficiently
rich for the task at hand. Broadly speaking, one may consider
predictive or contrastive objectives. For predictive objectives, see
for example, \cite{ha2018world, hafner2020dream,
  schwarzer2021spr}. Contrastive objectives aim to leverage invariance
and discriminability, grouping "similar" observations together while
separating "dissimilar" ones. Examples include \cite{oord2018cpc,
  srinivas2020curl, stooke2021atc}.

Not only learning observations is important, but exploring the
diversity of those observations is also crucial. A widely adopted
strategy is to provide intrinsic rewards. For example, a player may
favor observations where internal estimations are failing to
accurately predict their targets. See, for example,
\cite{pathak2017icm,burda2019rnd,guo2022byolexplore}. Another approach
is to promote unfamiliar states to the player, such as,
\cite{badia2020ngu,badia2020agent57,saade2023recode}.

To conclude this brief overview, we point out that even when there is
only a single decision-maker, setting up an environment through a
Markov decision process (or one of its variants) does not capture the
full picture. Optimization is, of course, crucial for formalizing
which observations are preferred, and it provides strong guidance for
designing the players, as previously discussed. However, when many
players interact, seeking compatible strategies without accounting for
their internal estimations does not necessarily provide clear guidance
on which observations are going to be favored by the players. We
remark that, in Definition \ref{def:Uncertain-Equilibria} of uncertain
equilibrium, each player optimizes their own objective.

\subsection{A learning algorithm for CartPole}

We first briefly illustrate a method for learning CartPole. The goal
is not to propose a better learning method, but a novel one to draw an
attention to potential variety of learning algorithms for forming
estimations. We learn $\hat\phi$ with fixed horizon $T=8$. It is a
value-based method with model given. We trained value networks without
Bellman-type updates and instead promote such time-consistency
afterwards.

Let us revisit the basics of the CartPole problem. The state space is
$\dbS = \dbR^4$ representing position, velocity, angle, and angular
velocity. The action space is $\dbA = \{0,1\}$, where the actions
represent applying force to the left or right of the cart. The goal is
to keep the pole attached to the cart in the upright position.

The player has a memory, a function of observations, for recording
observations, including states, actions, and episode-level
evaluations. After each episode, a performance metric is computed as
the average of the agent’s relative and absolute performance, and
assigned as the evaluation of the episode. Here, the relative
performance is computed from the players moving average of how many
steps pole was upright. The absolute performance is computed depending
on the maximum potential steps, which is $500$.

In this problem, $\hat p$ is deterministic and can be
learned. However, as it is not of our interest, we took it as
given. We also set $\hat F = 0$ to only model the state values
$\hat\phi$. Introduce $\{\cN_\phi^k\}_{k=1}^{K_\phi}$ for some
$K_\phi\in\dbN$ as neural networks $\dbS \to \dbR_+$ for state
values. Let
$\fL_\phi: \sE\times\sM_\phi\to (\hat\O\times \dbS\to\dbR_+)$ and
\begin{equation*}
  \fL_\phi({}^n\cO,{}^{n-1}\hat\phi)
  := \sum_{k=1}^{K_\phi}\cN_\phi^k\ \ind{k}(\hat\o),
  \quad \hat\o \in \hat\O:= \{1,\dots,K_\phi\},\ \text{ and }
  \ \hat\dbP \text{ uniform}.
\end{equation*}
Here, ${}^n\hat\phi := \fL_\phi({}^n\cO,{}^{n-1}\hat\phi)$ is
representing abstractly the whole training process. $\hat\dbP$ is
taken as uniform means that the player has no information about which
network is providing better estimations, which is again for simplicity
only. We took the range of each $\cN_\phi^k$ as $[0,100]$, and trained
these networks relying on the memory. After $6$ episodes, the player
goes over the memory to recall best and worst performances, and forms
clusters of states to assign higher or lower values to them. We note
that, values are changing as the performance increase over time. Also,
the training is not done by assigning expected values, but instead
directly assigns higher or lower values. Then, we separately promoted
time-consistency of these assignments.

To ensure that the value networks are properly trained, we consider
the set of all controls $\dbA^T = \{0,1\}^T$ and index them by
$\{\a^k\}_{k=1}^{2^T}$.\footnote{We note that it is considerably
  easier learning controls in this simple setting, since $\dbA^T$ is
  finite. We first included a separate policy learning to generate a
  small amount of controls, and aimed to use value networks to select
  from them, but the player performs quite well even without properly
  trained value networks. Hence, we omitted from the discussion and
  evaluate all of the potential controls to make sure value is
  well-trained.} Then, we let
$\fL_\pi:\sM_\phi\to (\tilde\O\times\hat\O\times\dbS \to \dbA^T)$ as
\begin{equation*}
  \begin{aligned}
    &\fL_\pi := \sum_{k=1}^{K_\phi}\sum_{\ell=1}^{2^T} \delta_{\a^\ell}
    \ind{l}(\tilde\o)\ind{k}(\hat\o)\quad \text{where,}\ \ 
    \tilde\o\in\tilde\O := \{1,\dots, 2^T\},\ \ 
    \\&
    \hat\dbP(\tilde\o,\hat\o) := \frac{1}{Z}
    \exp\big(J(\hat \o, x, \a^{\tilde\o})\big),\ \  \text{and}\ \
    J(\cdot, x;\a) := \dbE^{\dbP^{x,\a}} \big[\hat\phi(\cdot,X_T)\big]
  \end{aligned}
\end{equation*}
Here, we extended the probability space by adding $\tilde\O$ to
separately keep track of randomness coming from value networks, and
the policy that they induce. $\hat\dbP$ is also now viewed as a
probability over $\tilde\O\times\hat\O$, and $Z$ is the normalizing
constant.

\begin{figure}[h]
  \begin{center}
    \includegraphics[width=1.0\textwidth]{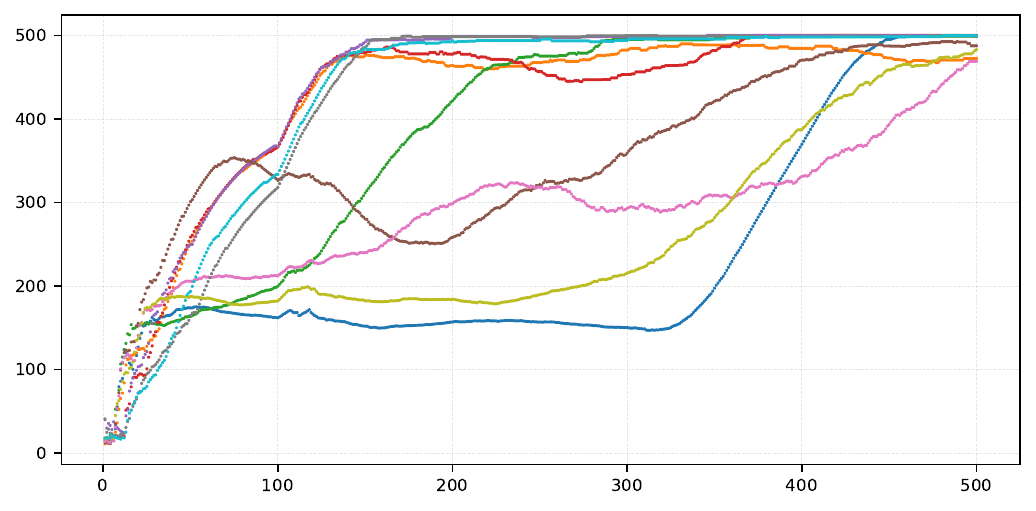}
    \caption{Performance of the CartPole Game Across 10 Selected
      Runs. The x-axis represents the number of games (or episodes),
      while the y-axis shows the total reward for each episode. Each
      line corresponds to one of the best 10 runs out of 32, with
      lines showing the moving average calculated from up to the last
      100 episodes.}
    \label{fig:CartPole}
  \end{center}
\end{figure}

Lastly, given $(\hat\phi, \hat\pi)$ as $(\fL_\phi, \fL_\pi)({}^n\cO)$,
the behavior
$\U:\sM_\phi\times \sM_{\pi}\to (\tilde\O\times\hat\O\times \dbS \to \dbA)$ is
taken as
\begin{equation*}
  \begin{aligned}
    \U(\hat\pi)(\hat\o,x) &:=
    \sum_{k=1}^{K_\phi}\sum_{\ell=1}^{2^T} \delta_{\a^\ell(x)}
    \ind{l}(\tilde\o)\ind{k}(\hat\o)
  \end{aligned}
\end{equation*}
In fact, the behavior commits for $T$ steps, but that would need to
introduce the parameter $t$ and we omitted it for brevity. See Figure
\ref{fig:CartPole} for the performance, and refer to the repository
\cite{CartPoleToyModel} for implementation details. Automatically
generated reports, produced using LLMs, are also provided to enhance
the accessibility of the implementation.

\subsection{Multi-agent Reinforcement Learning}

Our aim in this section is to provide a broad view of the multi-agent
reinforcement learning literature through the lens of the language
developed in this work. Once multiple players are present, relevant
considerations expand dramatically. Agents may model opponents,
communicate, teach, manipulate, or form temporary alliances, and it is
difficult to place principled bounds on such possibilities. In
particular, if a collection of agents is capable of developing a
language to communicate observations or situations they have
encountered, this opens the door to higher-order interactions: agents
may revisit past experiences, reassess estimations, propose
alternative plans, negotiate over future actions, implement voting
mechanisms, or assign specialized roles within the group. While such
behaviors are routine for us, they illustrate how rich multi-agent
learning can become once agents are viewed as structured
decision-makers rather than as static policies.

We begin by mentioning a major line of work in multi-agent learning,
which is motivated by convergence to equilibrium via regret
minimization. Initiated by \cite{ZJBP2007} in games with incomplete
information, these methods iteratively update policies so as to
minimize regret, thereby converging to unexploitable strategies. This
approach has led to notable success in heads-up limit Hold’em poker
\cite{BBJT2015}. Subsequent work has focused on scaling beyond tabular
representations, including neural network approximations
\cite{BLGS2019} and their integration with search-based planning
\cite{BBLG2020}. This line of research is fundamentally
equilibrium-oriented. From the perspective of a central designer
seeking stable strategies to announce, such methods play a crucial
role. However, as our focus is on player design rather than
equilibrium computation, we do not pursue this direction further. We
refer the interested reader to earlier equilibrium-motivated
approaches based on Q-learning \cite{HW2003}, and to \cite{BRMFTG2018}
for an analysis of convergence dynamics in differentiable games.

To extend value-based designs to multi-agent settings, one direct
approach is to treat each player as unaware of the others and to
regard the remaining players as an unknown component of the
environment. In this case, each player $(\cO^i, \fL_Q^i, \U^i)$ still
relies on a single type of estimate, such as
$\sM_Q^i := \{\dbS \times \dbA^i \to \dbR\}$, together with a common
behavior $\U^i$. However, unlike the single-agent setting, the
optimization target of the learning algorithms is no longer fixed by
the environment, since the behaviors of other players are themselves
evolving.\footnote{For simplicity, we write estimates as functions of
  the state $\dbS$. In practice, observations are often processed
  through additional estimates, such as recurrent neural networks, and
  the output space of those estimates is used instead. We omit such
  considerations to focus on what is new in the multi-agent case.}

Without allowing players to explicitly acquire additional estimates
about their opponents, one can still couple them through a shared
estimate. A large body of the literature studies this
centralized-training–decentralized-execution regime, primarily in
cooperative settings. Since the learning algorithm is centralized,
the training phase is more naturally viewed as a single player
$(\vec\cO, \fL_Q, \vec \U)$ with high-dimensional observations and
behaviors.

Consider, for example, a shared estimate space
$\sM_Q := \{\dbS \times \vec\dbA \to \dbR\}$ with learning algorithm
$\fL_Q: \vec \sE\times \vec\sM_Q\times \sM_Q \to \vec\sM_Q\times
\sM_Q$. As long as $\U^i$ depends only on $\sM_Q^i$, once training is
complete, one can define individual players as $(\cO^i, \fL^i, \U^i)$,
where $\fL^i$ yields $Q^i$ as a constant in order to freeze training.

There are several ways to realize such a single-player formulation.
For instance, in \cite{sunehag2018value}, the authors define $Q$ as
the sum of the individual $Q^i$'s. In \cite{rashid2018qmix}, this
approach is generalized by introducing an additional estimate
$\varphi: \vec\sM_Q \to \sM_Q$ satisfying
$\partial_{Q^i} \varphi \geq 0$, ensuring that the learned estimates
$Q^i$ respect each player's local maximization behavior. In
\cite{son2019qtran}, the authors further generalize this framework by
relaxing the monotonicity constraint and introducing consistency
conditions that allow a broader class of decompositions while
preserving individual players' behavior. Finally, in
\cite{wang2021qplex}, the analysis is carried out in the advantage
space by introducing two estimates $V$ and $A$ to define $Q$, while
maintaining a general representation.

The same centralization idea can also be applied to actor--critic
designs. In this case, the planned behavior
${}^n\vec\U \in \vec\sM_\U$ of the central (single) agent is fixed
after training, and the individual agents are equipped with mappings
$\U^i$ that constantly yield ${}^n\U^i \in \sM_\U^i$.

In \cite{lowe2017multi}, the authors apply a deterministic policy
gradient method with a central player $(\vec\cO, \vec\fL_Q, \vec\U)$,
where
\begin{equation*}
  \begin{aligned}
    &\fL_Q^i: \vec\sE\times \vec\sM_Q\times \vec\U \to \sM_Q^i,
    \\&\U^i: \sE^i\times \sM_Q^i \times \sM_\U^i \to \sM_\U^i,
  \end{aligned}
\end{equation*}
and $\sM_\U^i = \{\dbS^i \to \dbA^i\}$,
$\sM_Q^i = \{\vec\dbS \times \vec\dbA \to \dbR\}$. In
\cite{foerster2018counterfactual}, the authors consider stochastic
policy gradients, and hence
$\sM_\U^i = \{\O^u \times \dbS^i \to \dbA^i\}$ when written as a
random variable. They introduce an advantage estimate that compares
the current $Q$ estimate to an average taken over the planned behavior
${}^n\U^i$, in order to provide individual credit assignment. In
\cite{yu2022surprising}, the authors demonstrate the effectiveness of
proximal policy optimization in this regime. This core method has also
been successfully used in large-scale systems such as OpenAI Five for
\emph{Dota 2} \cite{berner2019dota}.

In this centralized regime, communication has also been studied
extensively in the literature. In this context, communication often
refers to the introduction of two additional estimates. First, each
agent constructs messages from its individual observations, taking
values in a Euclidean space, $C\in \sM_C^i := \{\sE^i \to
\dbR^k\}$. Second, each player maintains an estimate that aggregates
the incoming messages,
$D\in \sM_D^i := \{\vec \dbR^k \to \dbR^{k'}\}$. Within this
framework, the learning algorithm of the central player can, for
example, be formulated as
\begin{equation*}
  \fL_Q :
  \vec\sE \times \vec\sM_C \times \vec\sM_D \times \vec\sM_Q\times \sM_Q
  \to \vec\sM_C \times \vec\sM_D \times \vec\sM_Q\times \sM_Q
\end{equation*}
allowing the estimates $C$ and $D$ to be updated jointly with the
value estimates, for instance via backpropagation. The value functions
may then incorporate the aggregated messages in their domain,
$\sM_Q^i := \{ \dbS \times \dbA^i \times \dbR^{k'} \to \dbR\}$, or,
alternatively, the planned behavior itself may be defined to depend
explicitly on the aggregated messages.

In \cite{foerster2016learning}, the authors implement communication
both in independent Q-learning and in a centralized variant, following
the general structure described above. In \cite{das2019tarmac}, the
authors augment messages with a notion of directionality, allowing
each agent to attend to incoming messages along a chosen direction. In
\cite{sukhbaatar2016learning}, the authors adopt a centralized
single-agent perspective, and introduce multiple communication layers
$D^1, D^2, D^3, \dots$, in which messages are iteratively
averaged. This architecture enables agents to communicate at higher
levels, even when they are not directly connected. In
\cite{singh2019learning}, the authors extend this line of work to
settings with individual agents, including competitive scenarios, and
introduce a gating mechanism that allows agents to learn when to
exchange messages and when to remain silent. In \cite{jiang2020graph},
attention mechanisms are used to aggregate messages across multiple
communication layers. Finally, in \cite{ryu2021multi}, the
attention-based framework is further extended by incorporating a
gating mechanism through a learnable adjacency matrix, treating the
communication graph itself as a differentiable estimate.

A large-scale project worth mentioning in the context of communication
is \cite{FAIR2022Diplomacy}, which studies the turn-based game
\emph{Diplomacy}, a setting that allows natural-language communication
between players. The authors train a large language model on human
gameplay data to generate dialogue and introduce explicit estimates of
opponents’ intents. Policies are then learned in a manner grounded in
human play, integrating strategic decision-making with models of
communication and inferred opponent intentions. In effect, this
approach constructs a richly structured model of a human player. This
example further illustrates that such games cannot be adequately
captured by standard external formulations.

We remark that the above example already includes elements of opponent
modeling. Let us mention a few important works on opponent modeling.
Depending on the structure of the game, it may suffice for a player to
observe how other agents influence the state of the game. For example,
in another large-scale project \cite{vinyals2019grandmaster}, the
authors train independent agents for the game \emph{StarCraft II},
where players compete through carefully designed self-play as well as
against opponents constructed to exploit their strategies, with
policies guided in part by human gameplay data. However, a more
sophisticated player may need to maintain explicit estimates dedicated
to its opponents. In \cite{he2016opponent}, the authors introduce an
additional estimate that constructs an embedding of the opponent based
on observations.  First, they extend the $Q$-function to incorporate
this embedding as an additional input. Second, they consider a
collection of $Q$-functions and use the learned embedding to assign
weights to them. In these approaches, the learning algorithm for the
opponent embedding is initially embedded within the learning procedure
for the standard $Q$-estimate. Finally, the authors propose
introducing a separate learning algorithm specifically for the
opponent embedding in order to enhance its expressiveness. In
\cite{rabinowitz2018machine}, rather than constructing a full model of
a player, the authors focus on learning a particular estimate for
characterizing an observed opponent. They introduce one estimate that
aggregates information across observations from multiple episodes of
the observed player, and a second estimate that depends on the first
while incorporating partial observations within a single episode.
Together with the current observation, the outputs of these two
estimates are then used as inputs to a third estimate, whose learning
objective is to predict properties of the opponent being observed. In
\cite{jaques2019social}, the authors introduce intrinsic rewards for
influencing the behavior of other agents, and demonstrate that
accounting for such influence can promote coordination and
communication. They further show that these intrinsic rewards can be
trained in a decentralized manner, provided that agents acquire
predictive estimates of other agents’ action distributions.

We have aimed to emphasize the importance of player design. The
literature is extensive and spans many different directions. To
illustrate how broad the design space can be, we conclude by pointing
to recent works in which players incorporate large language models as
part of their estimates. For example, \cite{park2023generative}
demonstrates the emergence of believable human behavior, while
\cite{qian-etal-2024-chatdev,hong2024metagpt} introduce teams of
agents designed for collaborative software development.

\vspace{-0.5em}
\section{Conclusion}

We have reframed the problem of decision-making from the perspective
of the player and, in essence, abstracted the constructions in
reinforcement learning. Traditional approaches adopt a viewpoint
external to the players. Even when there is a single decision-maker,
the external setting may guide the design but does not suffice to
capture the required complexity. When many players interact,
considering only compatible strategies is not a sufficiently rich
objective, particularly for competing players.

Our broader vision is to emphasize the inherent complexity of
intelligent behavior. We do not act through a single estimate, but
through dynamically evolving layers of estimations and behaviors:
structures capable of working with any stream of observations,
adapting to novel situations, reconstructing, planning, and
predicting; forming diverse values both individually and
collectively. Understanding how a dynamical system achieves such
complexity remains far beyond our reach, but this work aims to provide
the foundational definitions for initiating a formal approach.

\bibliographystyle{ieeetr}
\bibliography{LearningApproachtoGames}% common bib file

@article{A1987,
  author    = {Robert J. Aumann},
  title     = {Correlated Equilibrium as an Expression of Bayesian Rationality},
  journal   = {Econometrica},
  year      = {1987},
  volume    = {55},
  number    = {1},
  pages     = {1--18},
  publisher = {Wiley on behalf of the Econometric Society},
  issn      = {0012-9682, 1468-0262},
  url       = {http://www.jstor.org/stable/1911154}
}

@incollection{B1974,
  author    = {H. D. Block},
  title     = {Random Orderings and Stochastic Theories of Responses (1960)},
  booktitle = {Economic Information, Decision, and Prediction: Selected Essays: Volume I, Part I Economics of Decision},
  year      = {1974},
  publisher = {Springer Netherlands},
  address   = {Dordrecht},
  pages     = {172--217},
  isbn      = {978-94-010-9276-0},
  doi       = {10.1007/978-94-010-9276-0\_8},
}

@incollection{B1951,
  author    = {George W. Brown},
  title     = {Iterative Solutions of Games by Fictitious Play},
  booktitle = {Activity Analysis of Production and Allocation},
  editor    = {Tjalling C. Koopmans},
  publisher = {Wiley},
  address   = {New York},
  year      = {1951},
  pages     = {374--376}
}

@book{FL1998,
  title     = {The Theory of Learning in Games},
  author    = {Fudenberg, Drew and Levine, David K.},
  publisher = {MIT Press},
  address   = {Cambridge, MA},
  series    = {Economic Learning and Social Evolution},
  volume    = {2},
  year      = {1998},
  isbn      = {9780262061940}
}

@book{L1959,
  author    = {Luce, R. Duncan},
  title     = {Individual Choice Behavior: A Theoretical Analysis},
  publisher = {John Wiley and Sons},
  address   = {New York},
  year      = {1959}
}

@book{MSZ2013,
  author    = {Maschler, Michael and Solan, Eilon and Zamir, Shmuel},
  title     = {Game Theory},
  publisher = {Cambridge University Press},
  address   = {Cambridge},
  year      = {2013},
  doi       = {10.1017/CBO9780511794216}
}

@incollection{M1974,
  author    = {McFadden, Daniel},
  title     = {Conditional Logit Analysis of Qualitative Choice Behavior},
  booktitle = {Frontiers in Econometrics},
  editor    = {Zarembka, Paul},
  publisher = {Academic Press},
  address   = {New York},
  year      = {1974},
  pages     = {105--142}
}

@article{N1951,
  ISSN = {0003486X, 19398980},
  URL = {http://www.jstor.org/stable/1969529},
  author = {John Nash},
  journal = {Annals of Mathematics},
  number = {2},
  pages = {286--295},
  publisher = {[Annals of Mathematics, Trustees of Princeton University on Behalf of the Annals of Mathematics, Mathematics Department, Princeton University]},
  title = {Non-Cooperative Games},
  urldate = {2025-11-08},
  volume = {54},
  year = {1951}
}

@book{NM1944,
  author    = {von Neumann, John and Morgenstern, Oskar},
  title     = {Theory of Games and Economic Behavior},
  publisher = {Princeton University Press},
  address   = {Princeton, NJ},
  year      = {1944}
}

@article{T1933,
  ISSN = {00063444},
  URL = {http://www.jstor.org/stable/2332286},
  author = {William R. Thompson},
  journal = {Biometrika},
  number = {3/4},
  pages = {285--294},
  publisher = {[Oxford University Press, Biometrika Trust]},
  title = {On the Likelihood that One Unknown Probability Exceeds Another in View of the Evidence of Two Samples},
  urldate = {2025-11-08},
  volume = {25},
  year = {1933}
}

@article{T1927,
  author  = {Thurstone, L. L.},
  title   = {A Law of Comparative Judgment},
  journal = {Psychological Review},
  year    = {1927},
  volume  = {34},
  number  = {4},
  pages   = {273--286},
  doi     = {10.1037/h0070288}
}

@article{ACF2002,
  author  = {Auer, Peter and Cesa-Bianchi, Nicol{\`o} and Fischer, Paul},
  title   = {Finite-Time Analysis of the Multi-Armed Bandit Problem},
  journal = {Machine Learning},
  year    = {2002},
  volume  = {47},
  number  = {2--3},
  pages   = {235--256},
  doi     = {10.1023/A:1013689704352}
}

@article{CCDP2020,
  author  = {Emilio Calvano and Giacomo Calzolari and Vincenzo Denicol{\`o} and Sergio Pastorello},
  title   = {Artificial Intelligence, Algorithmic Pricing, and Collusion},
  journal = {American Economic Review},
  year    = 2020,
  volume  = 110,
  number  = 10,
  pages   = {3267--3297},
  month   = {October},
  doi     = {10.1257/aer.20190623},
  url     = {https://www.aeaweb.org/articles?id=10.1257/aer.20190623}
}

@inproceedings{CL2011,
  title     = {An Empirical Evaluation of {T}hompson Sampling},
  author    = {Chapelle, Olivier and Li, Lihong},
  booktitle = {Advances in Neural Information Processing Systems},
  volume    = {24},
  pages     = {2249--2257},
  year      = {2011},
  editor    = {Shawe-Taylor, John and Zemel, Richard S. and Bartlett, Peter L. and Pereira, Fernando C. N. and Weinberger, Kilian Q.},
  publisher = {Curran Associates, Inc.},
  address   = {Red Hook, NY, USA}
}

@incollection{DFPPV2010,
  author    = {Constantinos Daskalakis and Rafael Frongillo and Christos H. Papadimitriou and George Pierrakos and Gregory Valiant},
  editor    = {Spyros Kontogiannis and Elias Koutsoupias and Paul G. Spirakis},
  title     = {On Learning Algorithms for Nash Equilibria},
  booktitle = {Algorithmic Game Theory},
  year      = {2010},
  publisher = {Springer Berlin Heidelberg},
  address   = {Berlin, Heidelberg},
  pages     = {114--125},
  isbn      = {978-3-642-16170-4},
  doi       = {10.1007/978-3-642-16170-4\_11},
  url       = {https://doi.org/10.1007/978-3-642-16170-4\_11}
}

@article{CH2017,
  author  = {Pierre Cardaliaguet and Saeed Hadikhanloo},
  title   = {Learning in Mean Field Games: The Fictitious Play},
  journal = {ESAIM: Control, Optimisation and Calculus of Variations},
  year    = {2017},
  volume  = {23},
  number  = {2},
  pages   = {569--591},
  doi     = {10.1051/cocv/2016004},
  url     = {https://doi.org/10.1051/cocv/2016004}
}

@book{CD2018a,
  title     = {Probabilistic Theory of Mean Field Games with Applications I: Mean Field {FBSDE}s, Control, and Games},
  author    = {Carmona, Ren{\'e} and Delarue, Fran{\c c}ois},
  publisher = {Springer},
  address   = {Cham},
  series    = {Probability Theory and Stochastic Modelling},
  volume    = {83},
  year      = {2018},
  doi       = {10.1007/978-3-319-58920-6}
}

@book{CD2018b,
  title     = {Probabilistic Theory of Mean Field Games with Applications II: Mean Field Games with Common Noise and Master Equations},
  author    = {Carmona, Ren{\'e} and Delarue, Fran{\c c}ois},
  publisher = {Springer},
  address   = {Cham},
  series    = {Probability Theory and Stochastic Modelling},
  volume    = {84},
  year      = {2018},
  doi       = {10.1007/978-3-319-58922-0}
}

@article{HMC2006,
  author  = {Minyi Huang and Roland P. Malham{\'e} and Peter E. Caines},
  title   = {Large-Population Stochastic Dynamic Games: Closed-Loop McKean--Vlasov Systems and the Nash Certainty Equivalence Principle},
  journal = {Communications in Information and Systems},
  year    = {2006},
  volume  = {6},
  number  = {3},
  pages   = {221--252},
  doi     = {10.4310/CIS.2006.v6.n3.a5},
  url     = {https://doi.org/10.4310/CIS.2006.v6.n3.a5}
}

@article{IZ2024,
  author  = {{\.{I}}{\c{s}}eri, Melih and Zhang, Jianfeng},
  title   = {Set Values for Mean Field Games},
  journal = {Transactions of the American Mathematical Society},
  year    = {2024},
  volume  = {377},
  number  = {10},
  pages   = {7117--7174},
  doi     = {10.1090/tran/9255}
}

@article{LL2007,
  author  = {Jean-Michel Lasry and Pierre-Louis Lions},
  title   = {Mean Field Games},
  journal = {Japanese Journal of Mathematics},
  year    = {2007},
  volume  = {2},
  number  = {1},
  pages   = {229--260},
  doi     = {10.1007/s11537-007-0657-8},
  url     = {https://doi.org/10.1007/s11537-007-0657-8},
  issn    = {1861-3624}
}

@unpublished{BIM2025,
  author = {Erhan Bayraktar and Melih {\.{I}}{\c{s}}eri and Neil
                  Mascarenhas},
  title  = {Algorithmic Collusion of Strategic Firms},
  note   = {Work in progress},
  year   = {2025}
}

@misc{TwoPlayerGame,
  author = {Melih {\.{I}}{\c{s}}eri},
  title  = {Two Player Game},
  howpublished = {GitHub repository},
  year   = {2025},
  note   = {Available at: \url{https://github.com/melihiseri/TwoPlayerGame}}
}

@misc{CartPoleToyModel,
  author = {Melih {\.{I}}{\c{s}}eri},
  title  = {CartPole Toy Model},
  howpublished = {GitHub repository},
  year   = {2025},
  note   = {Available at: \url{https://github.com/melihiseri/CartPole_ToyModel}}
}

@book{T2009,
  author    = {Kenneth E. Train},
  title     = {Discrete Choice Methods with Simulation},
  edition   = {2},
  publisher = {Cambridge University Press},
  address   = {Cambridge},
  year      = {2009}
}

@article{HM2000,
  author  = {Sergiu Hart and Andreu Mas-Colell},
  title   = {A Simple Adaptive Procedure Leading to Correlated Equilibrium},
  journal = {Econometrica},
  year    = {2000},
  volume  = {68},
  number  = {5},
  pages   = {1127--1150},
  doi     = {10.1111/1468-0262.00153},
  url     = {https://onlinelibrary.wiley.com/doi/10.1111/1468-0262.00153}
}

@article{Mnih2015,
  author  = {Volodymyr Mnih and Koray Kavukcuoglu and David Silver and Andrei A. Rusu and Joel Veness and Marc G. Bellemare and Alex Graves and Martin Riedmiller and Andreas K. Fidjeland and Georg Ostrovski and Stig Petersen and Charles Beattie and Amir Sadik and Ioannis Antonoglou and Helen King and Dharshan Kumaran and Daan Wierstra and Shane Legg and Demis Hassabis},
  title   = {Human-Level Control through Deep Reinforcement Learning},
  journal = {Nature},
  year    = {2015},
  volume  = {518},
  number  = {7540},
  pages   = {529--533},
  doi     = {10.1038/nature14236},
  url     = {https://doi.org/10.1038/nature14236},
  issn    = {1476-4687}
}

@inproceedings{vieillard2020munchausen,
  title     = {Munchausen Reinforcement Learning},
  author    = {Vieillard, Nino and Pietquin, Olivier and Geist, Matthieu},
  booktitle = {Advances in Neural Information Processing Systems},
  volume    = {33},
  pages     = {4235--4246},
  year      = {2020},
  editor    = {Larochelle, H. and Ranzato, M. and Hadsell, R. and Balcan, M.F. and Lin, H.},
  publisher = {Curran Associates, Inc.},
  address   = {Red Hook, NY, USA},
}

@inproceedings{vanhasselt2016deep,
  author    = {Hado van Hasselt and Arthur Guez and David Silver},
  title     = {Deep Reinforcement Learning with Double Q-Learning},
  booktitle = {Proceedings of the Thirtieth AAAI Conference on Artificial Intelligence (AAAI'16)},
  year      = {2016},
  publisher = {AAAI Press},
  pages     = {2094--2100},
  address   = {Phoenix, Arizona},
  url       = {https://ojs.aaai.org/index.php/AAAI/article/view/10295}
}

@inproceedings{fujimoto2018addressing,
  title     = {Addressing Function Approximation Error in Actor-Critic Methods},
  author    = {Fujimoto, Scott and van Hoof, Herke and Meger, David},
  booktitle = {Proceedings of the 35th International Conference on Machine Learning (ICML)},
  volume    = {80},
  pages     = {1587--1596},
  year      = {2018},
  publisher = {PMLR},
  address   = {Stockholm, Sweden},
  url       = {http://proceedings.mlr.press/v80/fujimoto18a.html}
}

@inproceedings{haarnoja2018soft,
  title     = {Soft Actor-Critic: Off-Policy Maximum Entropy Deep Reinforcement Learning with a Stochastic Actor},
  author    = {Haarnoja, Tuomas and Zhou, Aurick and Abbeel, Pieter and Levine, Sergey},
  booktitle = {Proceedings of the 35th International Conference on Machine Learning (ICML)},
  volume    = 80,
  pages     = {1861--1870},
  year      = 2018,
  publisher = {PMLR},
  address   = {Stockholm, Sweden},
  url       = {http://proceedings.mlr.press/v80/haarnoja18b.html}
}

@article{williams1992simple,
  author  = {Ronald J. Williams},
  title   = {Simple Statistical Gradient-Following Algorithms for Connectionist Reinforcement Learning},
  journal = {Machine Learning},
  year    = {1992},
  volume  = {8},
  number  = {3},
  pages   = {229--256},
  doi     = {10.1007/BF00992696},
  url     = {https://doi.org/10.1007/BF00992696},
  issn    = {1573-0565}
}

@inproceedings{sutton2000policy,
  title     = {Policy Gradient Methods for Reinforcement Learning with Function Approximation},
  author    = {Sutton, Richard S. and McAllester, David A. and Singh, Satinder P. and Mansour, Yishay},
  booktitle = {Advances in Neural Information Processing Systems},
  volume    = {12},
  pages     = {1057--1063},
  year      = {2000},
  editor    = {Solla, S. A. and Leen, T. K. and M\"{u}ller, K.-R.},
  publisher = {MIT Press},
  address   = {Cambridge, MA},
}

@inproceedings{mnih2016asynchronous,
  author    = {Volodymyr Mnih and Adria Puigdomenech Badia and Mehdi Mirza and Alex Graves and Timothy Lillicrap and Tim Harley and David Silver and Koray Kavukcuoglu},
  title     = {Asynchronous Methods for Deep Reinforcement Learning},
  booktitle = {Proceedings of the 33rd International Conference on Machine Learning (ICML 2016)},
  editor    = {Maria Florina Balcan and Kilian Q. Weinberger},
  series    = {Proceedings of Machine Learning Research},
  volume    = 48,
  pages     = {1928--1937},
  year      = 2016,
  address   = {New York, NY, USA},
  publisher = {PMLR},
  url       = {https://proceedings.mlr.press/v48/mniha16.html}
}

@article{schulman2017proximal,
  title   = {Proximal Policy Optimization Algorithms},
  author  = {Schulman, John and Wolski, Filip and Dhariwal, Prafulla and Radford, Alec and Klimov, Oleg},
  journal = {arXiv preprint arXiv:1707.06347},
  year    = {2017},
  url     = {https://arxiv.org/abs/1707.06347}
}

@inproceedings{schulman2016high,
  title     = {High-Dimensional Continuous Control Using Generalized Advantage Estimation},
  author    = {Schulman, John and Moritz, Philipp and Levine, Sergey and Jordan, Michael I. and Abbeel, Pieter},
  booktitle = {International Conference on Learning Representations (ICLR)},
  year      = {2016},
  url       = {https://arxiv.org/abs/1506.02438}
}

@inproceedings{bellemare2017distributional,
  author    = {Marc G. Bellemare and Will Dabney and R{\'e}mi Munos},
  title     = {A Distributional Perspective on Reinforcement Learning},
  booktitle = {Proceedings of the 34th International Conference on Machine Learning (ICML 2017)},
  series    = {Proceedings of Machine Learning Research},
  volume    = {70},
  pages     = {449--458},
  year      = {2017},
  address   = {Sydney, NSW, Australia},
  publisher = {PMLR},
  url       = {https://proceedings.mlr.press/v70/bellemare17a.html}
}

@inproceedings{dabney2018distributional,
  author    = {Will Dabney and Mark Rowland and Marc G. Bellemare and R{\'e}mi Munos},
  title     = {Distributional Reinforcement Learning with Quantile Regression},
  booktitle = {Proceedings of the Thirty-Second AAAI Conference on Artificial Intelligence (AAAI-18)},
  year      = {2018},
  pages     = {2892--2901},
  address   = {New Orleans, Louisiana, USA},
  publisher = {AAAI Press},
  url       = {https://ojs.aaai.org/index.php/AAAI/article/view/11791}
}

@inproceedings{dabney2018implicit,
  author    = {Will Dabney and Georg Ostrovski and David Silver and R{\'e}mi Munos},
  title     = {Implicit Quantile Networks for Distributional Reinforcement Learning},
  booktitle = {Proceedings of the 35th International Conference on Machine Learning (ICML 2018)},
  editor    = {Jennifer Dy and Andreas Krause},
  series    = {Proceedings of Machine Learning Research},
  volume    = {80},
  pages     = {1096--1105},
  year      = {2018},
  address   = {Stockholmsmässan, Stockholm, Sweden},
  publisher = {PMLR},
  url       = {https://proceedings.mlr.press/v80/dabney18a.html}
}

@inproceedings{yang2019fully,
  author    = {Derek Yang and Li Zhao and Zichuan Lin and Tao Qin and Jiang Bian and Tie-Yan Liu},
  title     = {Fully Parameterized Quantile Function for Distributional Reinforcement Learning},
  booktitle = {Proceedings of the 33rd Conference on Neural Information Processing Systems (NeurIPS 2019)},
  year      = {2019},
  publisher = {Curran Associates, Inc.},
  address   = {Red Hook, NY, USA},
  pages     = {556--566},
}

@inproceedings{osband2016deep,
  title     = {Deep Exploration via Bootstrapped {DQN}},
  author    = {Osband, Ian and Blundell, Charles and Pritzel, Alexander and Van Roy, Benjamin},
  booktitle = {Advances in Neural Information Processing Systems},
  volume    = {29},
  pages     = {4026--4034},
  year      = {2016},
  publisher = {Curran Associates, Inc.},
  address   = {Red Hook, NY, USA},
}

@article{silver2016mastering,
  author  = {David Silver and Aja Huang and Chris J. Maddison and Arthur Guez and Laurent Sifre and George van den Driessche and Julian Schrittwieser and Ioannis Antonoglou and Veda Panneershelvam and Marc Lanctot and Sander Dieleman and Dominik Grewe and John Nham and Nal Kalchbrenner and Ilya Sutskever and Timothy Lillicrap and Madeleine Leach and Koray Kavukcuoglu and Thore Graepel and Demis Hassabis},
  title   = {Mastering the Game of Go with Deep Neural Networks and Tree Search},
  journal = {Nature},
  year    = {2016},
  volume  = {529},
  number  = {7587},
  pages   = {484--489},
  doi     = {10.1038/nature16961},
  url     = {https://doi.org/10.1038/nature16961},
  issn    = {1476-4687}
}

@article{silver2018general,
  author  = {David Silver and Thomas Hubert and Julian Schrittwieser and Ioannis Antonoglou and Matthew Lai and Arthur Guez and Marc Lanctot and Laurent Sifre and Dharshan Kumaran and Thore Graepel and Timothy Lillicrap and Karen Simonyan and Demis Hassabis},
  title   = {A General Reinforcement Learning Algorithm that Masters Chess, Shogi, and Go through Self-Play},
  journal = {Science},
  year    = {2018},
  volume  = {362},
  number  = {6419},
  pages   = {1140--1144},
  doi     = {10.1126/science.aar6404},
  url     = {https://www.science.org/doi/10.1126/science.aar6404},
  issn    = {0036-8075}
}

@article{schrittwieser2020mastering,
  author  = {Julian Schrittwieser and Ioannis Antonoglou and Thomas Hubert and Karen Simonyan and Laurent Sifre and Simon Schmitt and Arthur Guez and Edward Lockhart and Demis Hassabis and Thore Graepel and Timothy Lillicrap and David Silver},
  title   = {Mastering Atari, Go, Chess and Shogi by Planning with a Learned Model},
  journal = {Nature},
  year    = {2020},
  volume  = {588},
  number  = {7839},
  pages   = {604--609},
  doi     = {10.1038/s41586-020-03051-4},
  url     = {https://doi.org/10.1038/s41586-020-03051-4},
  issn    = {1476-4687}
}

@article{ha2018world,
  title={World models},
  author={Ha, David and Schmidhuber, J{\"u}rgen},
  journal={arXiv preprint arXiv:1803.10122},
  year={2018}
}

@inproceedings{hafner2019learning,
  author    = {Danijar Hafner and Timothy Lillicrap and Ian Fischer and Ruben Villegas and David Ha and Honglak Lee and James Davidson},
  title     = {Learning Latent Dynamics for Planning from Pixels},
  booktitle = {Proceedings of the 36th International Conference on Machine Learning (ICML 2019)},
  editor    = {Kamalika Chaudhuri and Ruslan Salakhutdinov},
  series    = {Proceedings of Machine Learning Research},
  volume    = {97},
  pages     = {2555--2565},
  year      = {2019},
  address   = {Long Beach, California, USA},
  publisher = {PMLR},
  url       = {https://proceedings.mlr.press/v97/hafner19a.html}
}

@inproceedings{hafner2020dream,
  author    = {Danijar Hafner and Timothy Lillicrap and Jimmy Ba and Mohammad Norouzi},
  title     = {Dream to Control: Learning Behaviors by Latent Imagination},
  booktitle = {Proceedings of the 8th International Conference on Learning Representations (ICLR 2020)},
  year      = {2020},
  url       = {https://openreview.net/forum?id=S1lOTC4tDS}
}

@inproceedings{nagabandi2018neural,
  author    = {Anusha Nagabandi and Gregory Kahn and Ronald S. Fearing and Sergey Levine},
  title     = {Neural Network Dynamics for Model-Based Deep Reinforcement Learning with Model-Free Fine-Tuning},
  booktitle = {Proceedings of the 2018 IEEE International Conference on Robotics and Automation (ICRA)},
  year      = {2018},
  pages     = {7559--7566},
  doi       = {10.1109/ICRA.2018.8463189},
  address   = {Brisbane, Australia},
  publisher = {IEEE},
}

@inproceedings{chua2018deep,
  title     = {Deep Reinforcement Learning in a Handful of Trials Using Probabilistic Dynamics Models},
  author    = {Chua, Kurtland and Calandra, Roberto and McAllister, Rowan and Levine, Sergey},
  booktitle = {Advances in Neural Information Processing Systems},
  volume    = {31},
  pages     = {4754--4765},
  year      = {2018},
  publisher = {Curran Associates, Inc.},
  address   = {Red Hook, NY, USA},
}

@inproceedings{bacon2017option,
  author    = {Pierre{-}Luc Bacon and Jean Harb and Doina Precup},
  title     = {The Option-Critic Architecture},
  booktitle = {Proceedings of the Thirty-First AAAI Conference on Artificial Intelligence (AAAI-17)},
  year      = {2017},
  pages     = {1726--1734},
  address   = {San Francisco, California, USA},
  publisher = {AAAI Press},
  url       = {https://ojs.aaai.org/index.php/AAAI/article/view/10895}
}

@inproceedings{nachum2018data,
  author    = {Ofir Nachum and Shixiang Gu and Honglak Lee and Sergey Levine},
  title     = {Data-Efficient Hierarchical Reinforcement Learning},
  booktitle = {Advances in Neural Information Processing Systems 31 (NeurIPS 2018)},
  pages     = {3307--3317},
  year      = {2018},
  address   = {Montr{\'e}al, Canada},
  publisher = {Curran Associates, Inc.}
}

@inproceedings{levy2019learning,
  author    = {Andrew Levy and George Konidaris and Robert Platt},
  title     = {Learning Multi-Level Hierarchies with Hindsight},
  booktitle = {Proceedings of the 36th International Conference on Machine Learning (ICML 2019)},
  editor    = {Kamalika Chaudhuri and Ruslan Salakhutdinov},
  series    = {Proceedings of Machine Learning Research},
  volume    = {97},
  pages     = {3846--3855},
  year      = {2019},
  address   = {Long Beach, California, USA},
  publisher = {PMLR},
  url       = {https://proceedings.mlr.press/v97/levy19a.html}
}

@inproceedings{schwarzer2021spr,
  author    = {Max Schwarzer and Ankesh Anand and Rishabh Goel and R. Devon Hjelm and Aaron Courville and Philip Bachman},
  title     = {Data-Efficient Reinforcement Learning with Self-Predictive Representations},
  booktitle = {Proceedings of the 9th International Conference on Learning Representations (ICLR 2021)},
  year      = {2021},
  url       = {https://openreview.net/forum?id=uCQfPZwRaU}
}

@article{oord2018cpc,
  author  = {Aaron van den Oord and Yazhe Li and Oriol Vinyals},
  title   = {Representation Learning with Contrastive Predictive Coding},
  journal = {arXiv preprint arXiv:1807.03748},
  year    = {2018},
  url     = {https://arxiv.org/abs/1807.03748}
}

@inproceedings{srinivas2020curl,
  author    = {Michael Laskin and Aravind Srinivas and Pieter Abbeel},
  title     = {CURL: Contrastive Unsupervised Representations for Reinforcement Learning},
  booktitle = {Proceedings of the 37th International Conference on Machine Learning (ICML 2020)},
  editor    = {Hal Daumé III and Aarti Singh},
  series    = {Proceedings of Machine Learning Research},
  volume    = {119},
  pages     = {5639--5650},
  year      = {2020},
  address   = {Virtual Event},
  publisher = {PMLR},
  url       = {https://proceedings.mlr.press/v119/laskin20a.html}
}

@inproceedings{stooke2021atc,
  author    = {Adam Stooke and Kimin Lee and Michael Laskin and Pieter Abbeel},
  title     = {Decoupling Representation Learning from Reinforcement Learning},
  booktitle = {Proceedings of the 38th International Conference on Machine Learning (ICML 2021)},
  editor    = {Marina Meila and Tong Zhang},
  series    = {Proceedings of Machine Learning Research},
  volume    = {139},
  pages     = {9870--9879},
  year      = {2021},
  address   = {Virtual Event},
  publisher = {PMLR},
  url       = {https://proceedings.mlr.press/v139/stooke21a.html}
}

@inproceedings{pathak2017icm,
  author    = {Deepak Pathak and Pulkit Agrawal and Alexei A. Efros and Trevor Darrell},
  title     = {Curiosity-Driven Exploration by Self-Supervised Prediction},
  booktitle = {Proceedings of the 34th International Conference on Machine Learning (ICML 2017)},
  editor    = {Doina Precup and Yee Whye Teh},
  series    = {Proceedings of Machine Learning Research},
  volume    = {70},
  pages     = {2778--2787},
  year      = {2017},
  address   = {Sydney, NSW, Australia},
  publisher = {PMLR},
  url       = {https://proceedings.mlr.press/v70/pathak17a.html}
}

@inproceedings{burda2019rnd,
  author    = {Yuri Burda and Harrison Edwards and Amos Storkey and Oleg Klimov},
  title     = {Exploration by Random Network Distillation},
  booktitle = {Proceedings of the 7th International Conference on Learning Representations (ICLR 2019)},
  year      = {2019},
  url       = {https://openreview.net/forum?id=H1lJJnR5Ym}
}

@inproceedings{badia2020ngu,
  author    = {Adri{\`a} Puigdom{\`e}nech Badia and Bilal Piot and Steven Kapturowski and Pablo Sprechmann and Alex Vitvitskyi and Zhaohan Daniel Guo and Charles Blundell},
  title     = {Never Give Up: Learning Directed Exploration Strategies},
  booktitle = {Proceedings of the 8th International Conference on Learning Representations (ICLR 2020)},
  year      = {2020},
  url       = {https://openreview.net/forum?id=Sye57xStvB}
}

@inproceedings{badia2020agent57,
  author    = {Adri{\`a} Puigdom{\`e}nech Badia and Bilal Piot and Steven Kapturowski and Pablo Sprechmann and Alex Vitvitskyi and Zhaohan Daniel Guo and Charles Blundell},
  title     = {Agent57: Outperforming the Atari Human Benchmark},
  booktitle = {Proceedings of the 37th International Conference on Machine Learning (ICML 2020)},
  editor    = {Hal Daumé III and Aarti Singh},
  series    = {Proceedings of Machine Learning Research},
  volume    = {119},
  pages     = {507--517},
  year      = {2020},
  address   = {Virtual Event},
  publisher = {PMLR},
  url       = {https://proceedings.mlr.press/v119/badia20a.html}
}

@inproceedings{guo2022byolexplore,
  title     = {{BYOL}-Explore: Exploration by Bootstrapped Prediction},
  author    = {Guo, Zhaohan Daniel and Thakoor, Shantanu and P{\^i}slar, Miruna and Pires, Bernardo Avila and Altch{\'e}, Florent and Tallec, Corentin and Saade, Alaa and Calandriello, Daniele and Grill, Jean-Bastien and Tang, Yunhao and Valko, Michal and Munos, R{\'e}mi and Azar, Mohammad Gheshlaghi and Piot, Bilal},
  booktitle = {Advances in Neural Information Processing Systems},
  volume    = {35},
  pages     = {36965--36978},
  year      = {2022},
  publisher = {Curran Associates, Inc.},
  address   = {Red Hook, NY, USA},
}

@article{saade2023recode,
  author  = {Alaa Saade and Steven Kapturowski and Daniele Calandriello and Charles Blundell and Pablo Sprechmann and Leopoldo Sarra and Oliver Groth and Michal Valko and Bilal Piot},
  title   = {Unlocking the Power of Representations in Long-term Novelty-based Exploration (RECODE)},
  journal = {arXiv preprint arXiv:2305.01521},
  year    = {2023},
  url     = {https://arxiv.org/abs/2305.01521}
}

@article{KL1982,
  author  = {Kadane, Joseph B. and Larkey, Patrick D.},
  title   = {Subjective Probability and the Theory of Games},
  journal = {Management Science},
  year    = 1982,
  volume  = 28,
  number  = 2,
  pages   = {113--120},
  publisher = {INFORMS},
  doi     = {10.1287/mnsc.28.2.113}
}

@article{SPG2007,
  author  = {Shoham, Yoav and Powers, Rob and Grenager, Trond},
  title   = {If multi-agent learning is the answer, what is the question?},
  journal = {Artificial Intelligence},
  year    = {2007},
  volume  = {171},
  number  = {7},
  pages   = {365--377},
  doi     = {10.1016/j.artint.2006.02.006}
}

@techreport{SPG2003,
  author      = {Shoham, Yoav and Powers, Rob and Grenager, Trond},
  title       = {Multi-Agent Reinforcement Learning: a critical survey},
  institution = {Computer Science Department, Stanford University},
  year        = {2003},
  type        = {Technical Report},
  note        = {Unpublished manuscript}
}

@article{WL2017,
  author  = {Wright, James R. and Leyton-Brown, Kevin},
  title   = {Predicting human behavior in unrepeated, simultaneous-move games},
  journal = {Games and Economic Behavior},
  year    = {2017},
  volume  = {106},
  pages   = {16--37},
  doi     = {10.1016/j.geb.2017.09.009}
}

@inproceedings{HWB2016,
  title     = {Deep Learning for Predicting Human Strategic Behavior},
  author    = {Hartford, Jason S. and Wright, James R. and Leyton-Brown, Kevin},
  booktitle = {Advances in Neural Information Processing Systems},
  volume    = 29,
  pages     = {2424--2432},
  year      = 2016,
  publisher = {Curran Associates, Inc.},
  address   = {Red Hook, NY, USA},
}

@inproceedings{ZJBP2007,
  title     = {Regret Minimization in Games with Incomplete Information},
  author    = {Zinkevich, Martin and Johanson, Michael and Bowling, Michael and Piccione, Carmelo},
  booktitle = {Advances in Neural Information Processing Systems},
  volume    = {20},
  pages     = {1729--1736},
  year      = {2007},
  publisher = {Curran Associates, Inc.},
  address   = {Red Hook, NY, USA},
}

@article{BBJT2015,
  author    = {Bowling, Michael and Burch, Neil and Johanson, Michael and Tammelin, Oskari},
  title     = {Heads-up Limit Hold'em Poker is Solved},
  journal   = {Science},
  year      = {2015},
  volume    = {347},
  number    = {6218},
  pages     = {145--149},
  doi       = {10.1126/science.1259433}
}

@inproceedings{BLGS2019,
  title     = {Deep Counterfactual Regret Minimization},
  author    = {Brown, Noam and Lerer, Adam and Gross, Sam and Sandholm, Tuomas},
  booktitle = {Proceedings of the 36th International Conference on Machine Learning (ICML)},
  volume    = {97},
  pages     = {793--802},
  year      = {2019},
  publisher = {PMLR},
  address   = {Long Beach, CA, USA},
  url       = {http://proceedings.mlr.press/v97/brown19b.html}
}

@inproceedings{BBLG2020,
  title     = {Combining Deep Reinforcement Learning and Search for Imperfect-Information Games},
  author    = {Brown, Noam and Bakhtin, Anton and Lerer, Adam and Gong, Qucheng},
  booktitle = {Advances in Neural Information Processing Systems},
  volume    = {33},
  pages     = {13638--13650},
  year      = {2020},
  publisher = {Curran Associates, Inc.},
  address   = {Red Hook, NY, USA},
}

@inproceedings{BRMFTG2018,
  title     = {The Mechanics of n-Player Differentiable Games},
  author    = {Balduzzi, David and Racani{\`e}re, S{\'e}bastien and Martens, James and Foerster, Jakob and Tuyls, Karl and Graepel, Thore},
  booktitle = {Proceedings of the 35th International Conference on Machine Learning (ICML)},
  volume    = {80},
  pages     = {354--363},
  year      = {2018},
  publisher = {PMLR},
  address   = {Stockholm, Sweden},
  url       = {http://proceedings.mlr.press/v80/balduzzi18a.html}
}

@article{HW2003,
  author  = {Hu, Junling and Wellman, Michael P.},
  title   = {Nash Q-Learning for General-Sum Stochastic Games},
  journal = {Journal of Machine Learning Research},
  year    = {2003},
  volume  = {4},
  number  = {Nov},
  pages   = {1039--1069},
  url     = {https://www.jmlr.org/papers/v4/hu03a.html}
}

@inproceedings{sunehag2018value,
  title={Value-Decomposition Networks For Cooperative Multi-Agent Learning Based On Team Reward},
  author={Sunehag, Peter and Lever, Guy and Gruslys, Audrunas and Czarnecki, Wojciech Marian and Zambaldi, Vinicius and Jaderberg, Max and Lanctot, Marc and Sonnerat, Nicolas and Leibo, Joel Z and Tuyls, Karl and Graepel, Thore},
  booktitle={Proceedings of the 17th International Conference on Autonomous Agents and MultiAgent Systems},
  pages={2085--2087},
  year={2018}
}

@inproceedings{rashid2018qmix,
  title     = {QMIX: Monotonic Value Function Factorisation for Deep Multi-Agent Reinforcement Learning},
  author    = {Rashid, Tabish and Samvelyan, Mikayel and Schroeder, Christian and Farquhar, Gregory and Foerster, Jakob and Whiteson, Shimon},
  booktitle = {Proceedings of the 35th International Conference on Machine Learning (ICML)},
  volume    = {80},
  pages     = {4295--4304},
  year      = {2018},
  publisher = {PMLR},
  address   = {Stockholm, Sweden},
  url       = {http://proceedings.mlr.press/v80/rashid18a.html}
}

@inproceedings{son2019qtran,
  title     = {QTRAN: Learning to Factorize with Transformation for Cooperative Multi-Agent Reinforcement Learning},
  author    = {Son, Kyunghwan and Kim, Daewoo and Kang, Wan Ju and Hostallero, David Earl and Yi, Yung},
  booktitle = {Proceedings of the 36th International Conference on Machine Learning (ICML)},
  volume    = {97},
  pages     = {5887--5896},
  year      = {2019},
  publisher = {PMLR},
  address   = {Long Beach, CA, USA},
  url       = {http://proceedings.mlr.press/v97/son19a.html}
}

@inproceedings{wang2021qplex,
  title     = {QPLEX: Duplex Dueling Multi-Agent Q-Learning},
  author    = {Wang, Jianhao and Ren, Zhizhou and Liu, Terry and Yu, Yang and Zhang, Chongjie},
  booktitle = {International Conference on Learning Representations (ICLR)},
  year      = {2021},
  url       = {https://openreview.net/forum?id=Rcmk0xxIQV}
}

@inproceedings{lowe2017multi,
  title={Multi-Agent Actor-Critic for Mixed Cooperative-Competitive Environments},
  author={Lowe, Ryan and Wu, Yi and Tamar, Aviv and Harb, Jean and Abbeel, Pieter and Mordatch, Igor},
  booktitle={Advances in Neural Information Processing Systems},
  volume={30},
  year={2017}
}

@inproceedings{foerster2018counterfactual,
  title     = {Counterfactual Multi-Agent Policy Gradients},
  author    = {Foerster, Jakob and Farquhar, Gregory and Afouras, Triantafyllos and Nardelli, Nantas and Whiteson, Shimon},
  booktitle = {Proceedings of the AAAI Conference on Artificial Intelligence},
  volume    = {32},
  pages     = {2974--2982},
  year      = {2018},
  url       = {https://ojs.aaai.org/index.php/AAAI/article/view/11794}
}

@inproceedings{yu2022surprising,
  title={The Surprising Effectiveness of PPO in Cooperative Multi-Agent Games},
  author={Yu, Chao and Velu, Akash and Vinitsky, Eugene and Gao, Jiaxuan and Wang, Yu and Bayen, Alexandre and Wu, Yi},
  booktitle={Advances in Neural Information Processing Systems},
  volume={35},
  pages={24611--24624},
  year={2022}
}

@misc{berner2019dota,
      title={Dota 2 with Large Scale Deep Reinforcement Learning}, 
      author={OpenAI and Christopher Berner and Greg Brockman and
                  Brooke Chan and Vicki Cheung and Przemyslaw Debiak
                  and Christy Dennison and David Farhi and Quirin
                  Fischer and Shariq Hashme and Chris Hesse and Rafal
                  Józefowicz and Scott Gray and Catherine Olsson and
                  Jakub Pachocki and Michael Petrov and Henrique
                  P. d. O. Pinto and Jonathan Raiman and Tim Salimans
                  and Jeremy Schlatter and Jonas Schneider and Szymon
                  Sidor and Ilya Sutskever and Jie Tang and Filip
                  Wolski and Susan Zhang},
      year={2019},
      eprint={1912.06680},
      archivePrefix={arXiv},
      primaryClass={cs.LG},
      url={https://arxiv.org/abs/1912.06680}, 
}

@article{FAIR2022Diplomacy,
  author  = {{Meta Fundamental AI Research Diplomacy Team (FAIR)} and Bakhtin, Anton and Brown, Noam and Dinan, Emily and Farina, Gabriele and Flaherty, Colin and Fried, Daniel and Goff, Andrew and Gray, Jonathan and Hu, Hengyuan and Jacob, Athul Paul and Komeili, Mojtaba and Konath, Karthik and Kwon, Minae and Lerer, Adam and Lewis, Mike and Miller, Alexander H. and Mitts, Sasha and Renduchintala, Adithya and Roller, Stephen and Rowe, Dirk and Shi, Weiyan and Spisak, Joe and Wei, Alexander and Wu, David and Yates, Michael and Zhang, Huanyu and Zijlstra, Markus and Letychevsky, M. and others},
  title   = {Human-level play in the game of Diplomacy by combining language models with strategic reasoning},
  journal = {Science},
  year    = {2022},
  volume  = {378},
  number  = {6624},
  pages   = {1067--1074},
  doi     = {10.1126/science.ade9097},
  url     = {https://www.science.org/doi/10.1126/science.ade9097}
}

@inproceedings{foerster2016learning,
  title     = {Learning to Communicate with Deep Multi-Agent Reinforcement Learning},
  author    = {Foerster, Jakob N. and Assael, Yannis M. and de Freitas, Nando and Whiteson, Shimon},
  booktitle = {Advances in Neural Information Processing Systems},
  volume    = 29,
  pages     = {2137--2145},
  year      = 2016
}

@inproceedings{sukhbaatar2016learning,
  title     = {Learning Multiagent Communication with Backpropagation},
  author    = {Sukhbaatar, Sainbayar and Szlam, Arthur and Fergus, Rob},
  booktitle = {Advances in Neural Information Processing Systems},
  volume    = {29},
  pages     = {2244--2252},
  year      = {2016},
  publisher = {Curran Associates, Inc.},
  address   = {Red Hook, NY, USA}
}

@inproceedings{singh2019learning,
  title     = {Learning When to Communicate at Scale in Multiagent Cooperative and Competitive Tasks},
  author    = {Singh, Amanpreet and Jain, Tushar and Sukhbaatar, Sainbayar},
  booktitle = {International Conference on Learning Representations (ICLR)},
  year      = {2019},
  url       = {https://openreview.net/forum?id=rye7knCqK7}
}

@inproceedings{jiang2020graph,
  title={Graph Convolutional Reinforcement Learning},
  author={Jiang, Jiechuan and Dun, Chen and Huang, Tiejun and Lu, Zongqing},
  booktitle={International Conference on Learning Representations},
  year={2020}
}

@inproceedings{ryu2021multi,
  title={Multi-Agent Graph-Attention Communication and Teaming},
  author={Ryu, Jiechuan and Zhou, Hao and Park, Jongeui and Iosifidis, Alexandros},
  booktitle={Proceedings of the 20th International Conference on Autonomous Agents and MultiAgent Systems},
  pages={964--973},
  year={2021}
}

@inproceedings{das2019tarmac,
  title={TarMAC: Targeted Multi-Agent Communication},
  author={Das, Abhishek and Gervet, Th{\'e}ophile and Romoff, Joshua and Batra, Dhruv and Parikh, Devi and Rabbat, Michael and Pineau, Joelle},
  booktitle={Proceedings of the 36th International Conference on Machine Learning},
  pages={1538--1546},
  year=2019
}

@article{vinyals2019grandmaster,
  title={Grandmaster level in StarCraft II using multi-agent reinforcement learning},
  author={Vinyals, Oriol and Babuschkin, Igor and Czarnecki, Wojciech M and Mathieu, Micha{\"e}l and Dudzik, Andrew and Chung, Junyoung and Choi, David H and Powell, Richard and Ewalds, Timo and Georgiev, Petko and Oh, Junhyuk and Horgan, Dan and Kroiss, Manuel and Danihelka, Ivo and Huang, Aja and Sifre, Laurent and Cai, Trevor and Agapiou, John P and Jaderberg, Max and Vezhnevets, Alexander S and Leblond, R{\'e}mi and Pohlen, Tobias and Dalibard, Valentin and Budden, David and Sulsky, Yury and Molloy, James and Paine, Tom L and Gulcehre, Caglar and Wang, Ziyu and Pfaff, Tobias and Wu, Yuhuai and Ring, Roman and Yogatama, Dani and W{\"u}nsch, Dario and McKinney, Katrina and Smith, Oliver and Schaul, Tom and Lillicrap, Timothy and Kavukcuoglu, Koray and Hassabis, Demis and Apps, Chris and Silver, David},
  journal={Nature},
  volume={575},
  number={7782},
  pages={350--354},
  year={2019},
  doi={10.1038/s41586-019-1724-z},
}

@InProceedings{he2016opponent,
  title = 	 {Opponent Modeling in Deep Reinforcement Learning},
  author =       {He, He and Boyd-Graber, Jordan and Kwok, Kevin and Daum\'e, III, Hal},
  booktitle = 	 {Proceedings of The 33rd International Conference on Machine Learning},
  pages = 	 {1804--1813},
  year = 	 {2016},
  editor = 	 {Balcan, Maria Florina and Weinberger, Kilian Q.},
  volume = 	 {48},
  series = 	 {Proceedings of Machine Learning Research},
  address = 	 {New York, New York, USA},
  month = 	 {20--22 Jun},
  publisher =    {PMLR},
  url = 	 {https://proceedings.mlr.press/v48/he16.html},
}

@inproceedings{rabinowitz2018machine,
  title     = {Machine Theory of Mind},
  author    = {Rabinowitz, Neil and Perbet, Frank and Song, Francis and Zhang, Chiyuan and Eslami, S. M. Ali and Botvinick, Matthew},
  booktitle = {Proceedings of the 35th International Conference on Machine Learning (ICML)},
  volume    = {80},
  pages     = {4218--4227},
  year      = {2018},
  editor    = {Dy, Jennifer and Krause, Andreas},
  publisher = {PMLR},
  address   = {Stockholm, Sweden},
  series    = {Proceedings of Machine Learning Research},
  month     = {10--15 Jul},
  url       = {https://proceedings.mlr.press/v80/rabinowitz18a.html}
}

@inproceedings{jaques2019social,
  title     = {Social Influence as Intrinsic Motivation for Multi-Agent Deep Reinforcement Learning},
  author    = {Jaques, Natasha and Lazaridou, Angeliki and Hughes, Edward and Gulcehre, Caglar and Ortega, Pedro and Strouse, D.J. and Leibo, Joel Z. and de Freitas, Nando},
  booktitle = {Proceedings of the 36th International Conference on Machine Learning (ICML)},
  volume    = {97},
  pages     = {3040--3049},
  year      = {2019},
  publisher = {PMLR},
  address   = {Long Beach, CA, USA},
  url       = {http://proceedings.mlr.press/v97/jaques19a.html}
}

@inproceedings{qian-etal-2024-chatdev,
    title = "{C}hat{D}ev: Communicative Agents for Software Development",
    author = "Qian, Chen  and
      Liu, Wei  and
      Liu, Hongzhang  and
      Chen, Nuo  and
      Dang, Yufan  and
      Li, Jiahao  and
      Yang, Cheng  and
      Chen, Weize  and
      Su, Yusheng  and
      Cong, Xin  and
      Xu, Juyuan  and
      Li, Dahai  and
      Liu, Zhiyuan  and
      Sun, Maosong",
    booktitle = "Proceedings of the 62nd Annual Meeting of the Association for Computational Linguistics (Volume 1: Long Papers)",
    month = aug,
    year = "2024",
    address = "Bangkok, Thailand",
    publisher = "Association for Computational Linguistics",
    doi = "10.18653/v1/2024.acl-long.810",
    pages = "15174--15186",
}

@inproceedings{
hong2024metagpt,
title={Meta{GPT}: Meta Programming for A Multi-Agent Collaborative Framework},
author={Sirui Hong and Mingchen Zhuge and Jonathan Chen and Xiawu Zheng and Yuheng Cheng and Jinlin Wang and Ceyao Zhang and Zili Wang and Steven Ka Shing Yau and Zijuan Lin and Liyang Zhou and Chenyu Ran and Lingfeng Xiao and Chenglin Wu and J{\"u}rgen Schmidhuber},
booktitle={The Twelfth International Conference on Learning Representations},
year={2024},
url={https://openreview.net/forum?id=VtmBAGCN7o}
}

@inproceedings{park2023generative,
  title={Generative Agents: Interactive Simulacra of Human Behavior},
  author={Park, Joon Sung and O'Brien, Joseph C. and Cai, Carrie J. and Morris, Meredith Ringel and Liang, Percy and Bernstein, Michael S.},
  booktitle={Proceedings of the 36th Annual ACM Symposium on User Interface Software and Technology},
  pages={1--22},
  year={2023},
  publisher={Association for Computing Machinery},
  address={New York, NY, USA},
  doi={10.1145/3586183.3606763},
}

\end{document}